\documentclass{emulateapj}
\usepackage{graphicx}

\newcommand{\beq}	{\begin{equation}}
\newcommand{\eeq}	{\end{equation}}
\newcommand{\beqa}{\begin{eqnarray}}
\newcommand{\eeqa}{\end{eqnarray}}
\newcommand{\avg}[1]  {{\langle #1 \rangle}}

\def\simlt{\lower.5ex\hbox{$\; \buildrel < \over \sim \;$}}
\def\smgt{\lower.5ex\hbox{$\;. \buildrel > \over \sim \;$}}

\font\tenbi=cmmib10 
\newfam\bifam  \textfont\bifam=\tenbi

\font\tenbr=cmbx10
\newfam\brfam  \textfont\brfam=\tenbr

\font\squinttenbi=cmbx10 at 9pt
\scriptfont\brfam=\squinttenbi

\def\vecnabla{
              \setbox1=\hbox{$\bigtriangledown$}
                           \raise.45ex\hbox{$\bigtriangledown$\hskip-.97\wd1
                           $\bigtriangledown$\hskip-.97\wd1
                           $\bigtriangledown$\hskip-.97\wd1}
                           \raise.47ex\hbox{$\bigtriangledown$}}

\def\rsun{\ifmmode {\rm R}_{\mathord\odot}\else $R_{\mathord\odot}$\fi}
\def\msun{\ifmmode {\rm M}_{\mathord\odot}\else $M_{\mathord\odot}$\fi}
\def\lsun{\ifmmode {\rm L}_{\mathord\odot}\else $L_{\mathord\odot}$\fi}

\newcommand{\kms}	{{\rm km}\, {\rm s}^{-1}}

\def\tmb{\ifmmode {T_{\rm mb}^{13}(x,y,v)}\else $T_{\rm mb}^{13}(x,y,v)$\fi}

\begin{document}

\title{Impact of Winds from Intermediate-Mass Stars on Molecular Cloud Structure and Turbulence}



\author{Stella S.~R.~Offner}
\email{soffner@astro.umass.edu}
\affil{Department of Astronomy, University of Massachusetts, Amherst, MA 01002}

\author{H\'ector G. Arce}
\affil{Department of Astronomy, Yale University, New Haven, CT 06511}

\begin{abstract}
Observations of nearby molecular clouds detect ``shells", which are likely caused by winds from young main sequence stars. However, the progenitors of these observed features are not well characterized and the mass-loss rates inferred from the gas kinematics are several orders of magnitude greater than those predicted by atomic line-driven stellar wind models. We use magnetohydrodynamic simulations to model winds launching within turbulent molecular clouds and explore the impact of wind properties on cloud morphology and turbulence.  We find that winds do not produce clear features in turbulent statistics such as the Fourier spectra of density and momentum but do impact the Fourier velocity spectrum.  The density and velocity distribution functions, especially as probed by CO spectral lines, strongly indicate the presence and influence of winds. We show that stellar mass-loss rates for individual stars must be $\dot m_w \gtrsim 10^{-7} \msun$yr$^{-1}$, similar to those estimated from observations, to reproduce shell properties.  Consequently, we conclude that B and A-type main sequence stars have mass-loss rates several orders of magnitude larger that those predicted by models or that young stars are more variable than expected due to magnetic activity or accretion. 
\end{abstract}

\section{Introduction}
Turbulence within molecular clouds is thought to be an essential component of the star formation process, influencing cloud lifetimes, reducing star formation efficiencies and even setting the distribution of stellar masses \citep[][ and references therein]{MandO07, Offnerppvi}. However, the origin of molecular cloud turbulence and the means by which it is regenerated remain debated. One possibility is that young stars themselves inject sufficient energy to drive turbulent motions through protostellar outflows or radiatively driven winds, at least on parsec scales \citep[][and references therein]{frankppvi14}. Signatures of such activity exist within molecular clouds, however, it is difficult to quantify the true impact of these sources on the cloud motions.

Large-scale surveys of the Galactic plane, such as the GLIMPSE survey performed with the Spitzer Space Telescope, have discovered thousands of shell-like features in which molecular gas has been swept up and evacuated from large regions.  Many of the shells are directly associated with high-mass O-type stars, which produce ionizing photons as well as energetic radiatively driven winds. 
\citet{deharveng10} investigate a sub-sample of the Spitzer Galactic plane data and find that $\sim15$\% of the shells do not enclose ionized gas, which implies they may be driven by A or B-type stars.

Recently, high-resolution spectral observations of the nearby Perseus molecular cloud have identified a number of  shells (\citealt{arce11}, henceforth A11). The authors estimate that the momentum and energy in such features could account for approximately half of the total cloud turbulent energy,  much more than could be contributed by collimated outflows from protostars.  The larger shells are expanding with velocities in excess of the typical turbulent cloud velocity and create clear kinematic signatures in the gas. The shell regions influence the cloud velocity distribution significantly such that it deviates from the turbulent velocity distribution predicted by cloud simulations excluding feedback \citep{beaumont13}. 
Thus, it seems possible that the statistical properties of the cloud turbulence, which cascades from the largest scales, are appreciably different for turbulence influenced by internal stellar sources. For example, there is some evidence that turbulence metrics such as the momentum and energy Fourier spectra are different for protostellar outflow driven turbulence  (e.g., \citealt{swift08,carroll10}).

Within lower mass star forming molecular clouds like Perseus, O-type stars 
are not present. Instead, some of the shells appear to originate from less massive, but more numerous, B-type stars. Several such candidate progenitors are observed in and around the Perseus cloud (A11).  However, a number of the progenitors do not appear to have either associated O or B type stars and the shell origin is unknown.  \citet{nakamura12} find parsec-scale bubbles surrounding two small star clusters in the Orion region. They suggest that the bubbles are driven by the combined  winds from the most massive stars in these clusters, which are B and A-type stars. This could indicate that winds from multiple lower mass stars in a cluster may also inject significant energy to create such features.  The  B-type main sequence star $\rho$ Ophiuchi, which is located near the Ophiuchus star forming region, also has a sizable shell of 1.4pc \citep{wilking08}. 

If A and B-type stars generate some of these shells through radiatively driven winds, this presents a quandary for stellar atmospheric models. Mass-loss rates estimated from the shell masses, gas velocities, and Perseus cloud properties suggest winds with $\dot m_w \simeq 10^{-7}-10^{-6} \msun$ yr $^{-1}$. This is more than three orders of magnitude higher than those predicted by theoretical models \cite[][ see their Figure 3]{smith14}. In fact, around the transition mass between O and B-type stars,  radiative models suggest that the stellar radiation field may be too weak to drive mass-loss through optically thick spectral lines; this is known as the ``weak wind problem". Even without the complication of the weak wind problem, mass-loss rates for lower mass stars extrapolated from higher mass values would remain orders of magnitude too small (e.g.,\citealt{vink01}). 

 To date, little numerical work has been devoted to modelling winds from intermediate mass stars. Prior work has  focussed exclusively on winds from higher mass O-type stars and their impact on the gas in young clusters \citep{dale08,rogers13}. In these cases ionizing radiation plays a key role in creating bubbles, injecting momentum, and dispersing gas \citep{gendelev12,dale13,rogers14,geen15}. Only \citet{dale08} and \citet{dale13} have considered pure isotropic momentum driven winds. However, they only model mass-loss from stars with $m>10\msun$ and do not include magnetic fields. 

In this paper, we use hydromagnetic simulations to explore the range of mass-loss rates needed to reproduce the properties of observed shells in local star-forming clouds. We apply a variety of statistics to assess the degree to which the winds impact gas motions and assess whether winds impart detectable  signatures. We construct synthetic observations of the simulations in $^{12}$CO(1-0) to assess the accuracy of observational metrics in identifying and estimating properties of observed shells. We describe our simulations and models in \S \ref{sim_param}. We present results, including turbulent statistics and synthetic observations in \S\ref{results}. We discuss the results and evaluate alternative explanations in \S\ref{discussion} and summarize our conclusions in \S\ref{conclusions}.

\section{Simulation Parameters}\label{sim_param}

\subsection{Numerical Methods}\label{num_methods}

For our study we use the magneto-hydrodynamics adaptive mesh refinement (AMR) code {\sc orion} \citep[e.g.,][]{myers14}. {\sc orion} uses the {\sc chombo} toolset to solve partial differential questions on block-structured AMR meshes.  In the simulations we employ the constrained-transport magneto-hydrodynamic (CTMHD) module \citep{li12}, which is based on the finite volume scheme implemented in {\sc pluto} \citep{mignone07,mignone12}. 
{\sc orion} also has Lagrangian sink particles \citep{krumholz04}, which are inserted when collapsing gas exceeds the Jeans criterion on the maximum AMR level.  Extensions of the sink particle methodology allow {\sc orion}  to compute protostellar luminosities, launch protostellar bipolar outflows, and handle the accretion of magnetized gas \citep{,Offner09c,cunningham11,lee14}. In this paper we implement a new sink particle functionality: a sub-grid model that allows {\sc orion} to follow the launching of isotropic stellar winds. We present the wind methodology details in \S\ref{wind_model}.  {\sc orion} also has modules to compute self-gravity and radiative transfer via grey flux-limited diffusion \citep{miniati07, krumholzkmb07}, however, we do not include these in the present simulations.

	We treat the gas as an isothermal ideal gas such that its thermal energy is  $e = \rho k_B T / \left[ \mu(\gamma-1) \right]$,  where $\rho$ is the gas density, $k_B$ is the Boltzmann constant, $T$ is the gas temperature, $\mu$ is the mean particle mass, and $\gamma$ is the ratio of specific heats. We adopt $\gamma = 1.0001$ and $\mu = 3.9 \times 10^{-24}$g, which is the typical local value for cold molecular gas comprised of H$_2$ and He.


\begin{deluxetable*}{lcccccccc}
\tablecolumns{9}
\tablecaption{Model Properties\tablenotemark{a} \label{simprop}}
\tablehead{ \colhead{Model } &  
   \colhead{B($\mu G$)} &
  \colhead{ $\beta$} &
   \colhead{$t_i$ ($t_{\rm cross}$)} &
   \colhead{$t_{\rm run}$(Myr)} &
   \colhead{$\dot M_{\rm tot}(10^{-6 }\msun {\rm yr}^{-1})$\tablenotemark{b}} &
     \colhead{ $v_{\rm max}$ ($\kms$)} &
     \colhead{ $d_{x}$ (AU)} &
\colhead{Driving }}
\startdata
W1\_T1\_L0  & 13.5 &  0.1 & 1.6 & 0.15 & 41.7 & 200 &  4$\times 10^3$ & N \\  
W1\_T1  & 13.5 & 0.1 & 1.6 & 0.1 & 41.7 & 200 & 10$^3$ &N \\ 
W0\_T2  &   13.5 & 0.1 & 2.0 & 0.2 &  0 & 200 & 10$^3$  & N \\
W1\_T2 & 13.5 & 0.1 & 2.0 & 0.2 & 41.7 & 200 & 10$^3$ &N \\ 
W2\_T2 & 13.5 & 0.1 & 2.0 & 0.2 & 4.5 & 200 & 10$^3$ &Y \\  
W2\_T3\tablenotemark{c} &5.6 & 0.6 & 2.0 & 0.1 & 4.5 & 200 & 10$^3$ &Y \\  
W2\_T4\tablenotemark{c} &30.1 & 0.02 & 2.0 & 0.1 & 4.5 & 200 & 10$^3$ &N \\  
W3\_T2\tablenotemark{d} & 13.5 & 0.1 & 2.0 & 0.2 & $1\times10^{-3}$ & 10$^3$ & 10$^3$ &Y \\  
W4\_T2   & 13.5 & 0.1 & 2.0 & 0.15 & $1\times10^{-2}$ & 10$^3$ & 10$^3$ &Y   
\enddata
\tablenotetext{a}{Model name, initial mean magnetic field, ratio of thermal to magnetic pressure, the initial start time in crossing times, the run time, the total stellar mass-loss rate, the maximum wind velocity, minimum grid resolution, and whether external driving is continued simultaneously with winds or whether winds alone inject energy. All models have $L=5$pc, $M=3762 \msun$, $T_i=10$K and $N_*=5$.   All the calculations are first evolved with external driving but without sources to set up the initial turbulent conditions at $t=0$.} 
\tablenotetext{b}{The estimated mass-loss rate from all stellar winds in Perseus is $9.49 \times 10^{-6}\msun$yr$^{-1}$ (Arce et al.~2011). }
\tablenotetext{c}{Same wind model as W2\_T2 but a different initial magnetic field. }
\tablenotetext{d}{Same initial turbulence as run W2\_T2, but containing only a single isolated star with the given mass-loss rate.}
\end{deluxetable*}
\vspace{0.2in}

\begin{figure}
\epsscale{1.2}
\plotone{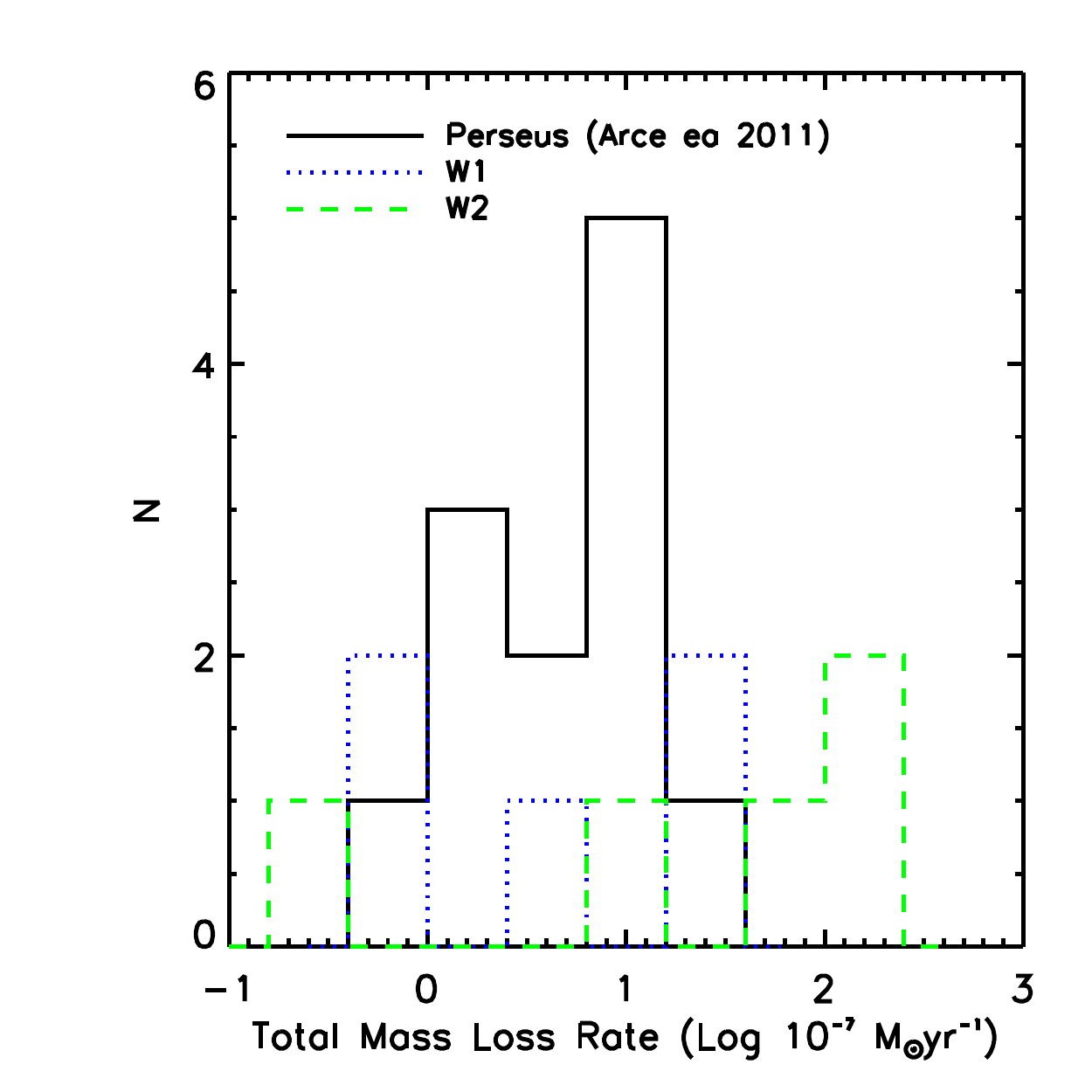}
\caption{Distribution of observed (black solid line) and simulated (colored lines) mass-loss rates. W1 and W2 indicate the mass-loss distributions for the models shown in Table \ref{simprop}. }\label{mass_loss_dist}
\vspace{0.1in}
\end{figure}

\subsection{Global Parameters}\label{global_param}

The simulations model a piece of a molecular cloud using a box with side $L=5$ pc and $M=3762~\msun$. The initial cloud temperature and Mach number are $T= 10$ K and $\mathcal{M}_{3D}=10.5$, respectively, which places the simulated cloud on the line-width size relation, $\sigma_{\rm 1D} = 0.72 R_{\rm pc}^{0.5}~ \kms$ \cite[e.g.,][]{MandO07}. To obtain initial density and velocity conditions, we ``drive" the simulation gas for approximately two mach crossing times without gravity by adding random large-scale perturbations to the velocity field \citep[e.g.][]{maclow99}. The perturbation wave numbers are  between 1 and 2, which corresponds to a linear size of $L/2-L$. 

The simulations start with an initially uniform vertical magnetic field, $B_z$, which becomes disordered over the turbulent driving period.  The simulations have plasma parameters (ratio of thermal pressure to magnetic pressure), $\beta=8 \pi \rho_0 c_s^2/B_z^2 = 0.02-0.6$, where $c_s$ is the sound speed and $\rho_0=2.04 \times 10^{-21}$g cm$^{-3}$ is the mean density. The fiducial Alf\'ven Mach number is $\mathcal{M}_{A,0} = \sqrt{12 \pi \rho_0} \sigma_{\rm 1D}/ B_0 = 2.3$.
The simulations follow four different wind launching models (W1-W4) and four different initial turbulent patterns (T1-T4),  which share the same Mach number but vary spatially and/or with bulk magnetic field strength at the initial time.
The simulation parameters are summarized in Table \ref{simprop}. 

In W1 and W2, we randomly place five stellar sources with different mass-loss rates. W1 has mass-loss rates ranging from $5.7\times 10^{-8}-2.0 \times 10^{-6}\msun$yr$^{-1}$ and W2 has mass-loss rates ranging from $2.6\times 10^{-8}-2.0 \times 10^{-6}\msun$yr$^{-1}$ (see Figure \ref{mass_loss_dist}). This number of sources is equivalent to the source number per volume identified in Perseus. The sources are meant to represent main sequence stars that have either formed in an earlier generation of stars or are older stars that formed elsewhere and wandered into the region. Either of these scenarios provides a plausible explanation for the observed sources in Perseus. For example, a 20 pc size cloud like Perseus has a dynamical time of $t_{\rm dyn} \simeq  L / \sigma_{1D} \sim 9$ Myr, which is sufficient time to form stars of 3-5 solar masses, which would reach the main sequence in $0.6-3$ Myr \citep[e.g.][]{wallerstein97}. Alternatively, the observed sources may have wandered into Perseus from a nearby region. B-type stars remain on the main sequence for approximately $1-5 \times 10^8$ years. If a star has a typical velocity of 1 $\kms$ \citep[e.g.,][]{foster15}, then it can travel about 1 parsec in 1 Myr, or a few hundred parsec before the end of its main sequence lifetime. This distance is sufficient such that some of the observed sources could have formed hundreds of millions of years earlier in a nearby cloud, whose gas is now dispersed.

In two additional simulations, W3 and W4, we examine the wind evolution for a single star with a prescribed mass-loss rate of $10^{-8}$ or $10^{-9} $ \msun yr$^{-1}$. These values are the mass-loss rates predicted for early O and late-type B stars, respectively, according to the \citet{vink01} model. Following the evolution of an isolated source avoids interactions between winds, which occurs in the other simulations and may boost the efficacy of gas entrainment and energy injection.

Finally, we also compute a purely turbulent model with no wind sources, W0. This model has the same initial turbulence as model T2, but the turbulence is simply allowed to decay without any energy input.

During the wind phase, we add additional refinement to the simulations according to the Truelove criterion with $N_J=0.125$ \citep{truelove97}. The application of a density gradient criterion, $\Delta \rho/\rho < 10$, ensures that wind shells are resolved by two additional levels of refinement for a maximum resolution of $\Delta x_2 = 10^3$ AU. We also rerun one simulation with no additional refinement to check the sensitivity of the results to the grid resolution (W1\_T1\_L0). In some of the models, once the stars are inserted and wind launching begins, external turbulent driving ceases and winds provide the only energy injection.  Whether driving is continued after the initialization phase is  indicated in Table \ref{simprop}.  In other cases, the models have both winds and external driving.

\subsection{Wind Model}\label{wind_model}

The three key properties of the wind model are the mass-loss rate, the wind velocity, and the wind temperature. Stellar mass is a factor only insofar as the mass-loss rate depends on it, otherwise, mass has no direct impact on the calculation since it has no dynamical impact without gravity. Star masses are randomly drawn such that their mass-loss rates are comparable to those inferred from the observed sources, i.e. $3\msun < m_* < 15 \msun$.
The most fundamental parameter of the simulation is the total mass-loss of the stars, $\dot M_{\rm tot}$. A random determination of 5 sources is inherently very stochastic, and a single object may dominate the total mass-loss budget.

\subsubsection{Mass-loss Rate}

The mass-loss rate for a given star can be related to fundamental parameters via an analytic fit to observed data as presented in \citet{vink01}. The fit depends on five stellar parameters: the stellar luminosity ($L_*$), stellar mass ($m_*$), effective temperature ($T_{\rm eff}$), metallicity ($Z$),  and $v_{\rm rat} = v_{\inf}/v_{\rm esc}$, which is the ratio of the terminal velocity to the escape speed at the surface. For $T_{\rm eff} < 2.25 \times 10^4$K:
\beqa
{\rm log} (\dot m) &=& -6.688 + 2.210 ~{\rm log}\left( \frac{L_*}{10^5 \lsun} \right) - 
             1.339{\rm ~log} \left( \frac{m_*}{30 \msun} \right) \nonumber \\
        & &     - 1.601 {\rm ~ log}\left( \frac{v_{\rm rat}}{2.0} \right) 
        + 1.07 {\rm ~log} \left( \frac{T_{\rm eff}}{2.0\times 10^4 {\rm K}} \right) \nonumber \\
        & & + 0.85 {\rm ~ log}(Z). 
\eeqa
For $T_{\rm eff} > 2.25 \times 10^4$K:
\beqa
{\rm log} (\dot m) &=& -6.697+2.194 {\rm ~ log} \left( \frac{L_*}{10^5\lsun} \right) 
 - 1.313 {\rm ~ log} \left( \frac{m_*}{30 \msun} \right) \nonumber \\
 & & -1.226 {\rm ~ log} \left( \frac{v_{\rm rat}}{2.0} \right)
+ 0.933 {\rm ~ log} \left( \frac{T_{\rm eff}}{4.0\times 10^4 {\rm K}} \right) \nonumber \\
& & - 10.92 \left[ {\rm ~ log}\left( \frac{T_{\rm eff}}{4.0\times 10^4 {\rm K}} \right) \right]^{2} + 0.85 \times {\rm ~ log}(Z).
\eeqa

From \citet{lamers95}:
\beqa 
v_{\rm rat} &=& 0.7 ~~~~~T_{\rm eff} < 1.25 \times 10^4 {\rm ~K}\\
v_{\rm rat} &=& 1.3 ~~~~~ 1.25 \times 10^4 < T_{\rm eff} < 2.1 \times 10^4 {\rm ~K}\\
v_{\rm rat} &=& 2.6 ~~~~~T_{\rm eff} > 2.1 \times 10^4 {\rm ~K}.
\eeqa
Note the values of the effective temperature for the transitions in $v_{\rm rat}$ are somewhat approximate, and different authors adopt slightly different values.

Although the stellar masses, and thus the mass-loss rates, are time dependent, the change is small over the simulation evolution, so the mass-loss rate can effectively be treated as a constant. Figure \ref{mass_loss_dist} shows the distribution of mass-loss rates for the observations and simulated sources.

\subsubsection{Wind Velocity}

The terminal wind velocity for OB line driven winds is a few hundred to a thousand $\kms$. From \citet{vink01}:
\beqa
{\rm log}(v) &=& 1.23 - 0.3 {\rm ~log}(L) + 0.55 {\rm ~log}(m) + 0.64{\rm ~log}(T_{\rm eff}) \nonumber \\
 & & + 0.13~ {\rm log}(Z). 
\eeqa
In the simulations, we limit the wind launching speed to some maximum velocity, $v_{\rm max}$ (see Table \ref{simprop}), in order to avoid inefficiently small time steps. Lower mass stars have smaller mass-loss rates and higher velocity winds. 

\subsubsection{Wind Temperature}

We assume that the wind is ionized, and thus set the launched gas temperature to $10^4$ K  and compute its thermal energy using $\mu = 2.13 \times 10^{-24}$g representing a mix of atomic H and He. We do not include cooling explicitly, such that hot wind gas can only cool by mixing with ambient cold material. Although lower mass stars like the Sun have coronae that are cooler, their winds are heated to much higher temperatures by convection \citep{owocki00}. Conversely, the winds of high-mass stars, which do not have a convective zone, are similar to that of their corona ($10^4-10^5$ K).

Inspection of the Perseus CO maps indicates that there is no CO inside most (but not all) of the bubbles. This suggests that the molecular gas has been dissociated and the temperatures inside the cavities are above $\sim 10^4$ K. H$^+$ is detected in one or two of the bubbles, namely the ones that appear to be driven by higher-mass stars. There is likely to be atomic hydrogen in some of the bubbles, but HI maps with sufficiently good resolution to confirm this directly do not exist. However, there is evidence of dust emission in some of the bubble interiors, which would require temperatures $T\lesssim 1100$ K, the dust sublimation temperature. 

The thermal pressure of the wind has little impact dynamically.
For example, if the thermal pressure of a gas with a average temperature, $T$, and density, $\rho$, is given by $P_{th}=\rho c_s^2$, where $c_s = \sqrt{k_BT/(\mu m_H)}$, then the fiducial simulation parameters yield an  average molecular gas pressure:
\beq
P/k_B = 5224 \left( \frac{\rho}{2.04\times 10^{-21}~{\rm g cm}^{-3}}\right) \left( \frac{T}{10~{\rm K}} \right) {\rm K cm}^{-3}.
\eeq
The pressure inside a wind bubble is the sum of the thermal pressure and gas ram pressure, where both depend upon the total mass-loss:
\beqa
P_{\rm wind} &=& P_{\rm ram} + P_{\rm th} \\
&=&  \frac{ \dot m_w t}{V} v_{\rm w}^2 + \frac{\dot m_w t}{V}c_{s,w}^2 ,
\eeqa 
where $V$ is the bubble volume. In a turbulent medium the evacuated region will not be exactly spherical, but we can still approximate $V \propto R^3$.
For a $10^4$K wind with terminal velocity $v=200~\kms$ the ratio of the thermal pressure to the ram pressure is $\simeq 1.5 \times 10^{-3}$  (see also \citealt{geen15}).


\begin{figure*}
\plotone{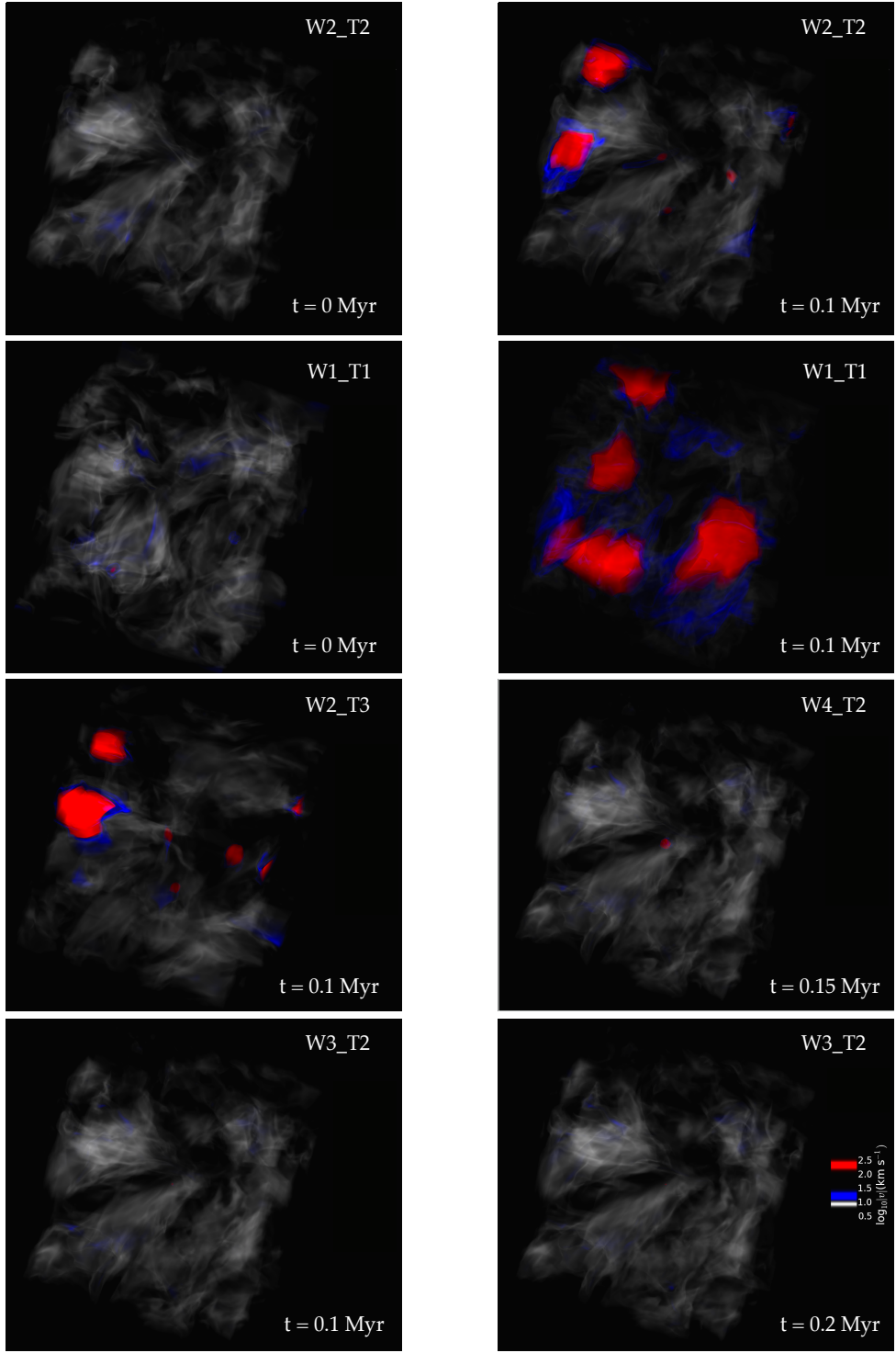}
\caption{Volume rendering of the gas velocity, where gray roughly traces the ambient cloud and red highlights the high-velocity wind regions.  Top: Run W2\_T2 ($\beta=0.1$) at times 0 Myr and 0.1 Myr (right). Top-Middle: Run W1\_T1 at times 0 Myr and 0.1 Myr (right).  Bottom-Middle: Run W2\_T3 ($\beta=0.6$) at 0.1 Myr and Run W4\_T2 ($\dot M=10^{-8}$ \msun yr$^{-1}$) at 0.15 Myr (right).  Bottom: Run W3\_T2 ($\dot M=10^{-9}$ \msun yr$^{-1}$) at times 0.1 Myr and 0.2 Myr (right).  Runs W2\_T2, W3\_T2 and W4\_T2 have the same gas distribution at $t=0$. For Runs W3\_T2 and W4\_T2 the wind source is in the center of the image.} \label{vol_rend}
\end{figure*}

 \subsection{Relation between Mass Loss and Shell Evolution}

We can use a simple calculation to relate the stellar mass-loss rate to the shell expansion.
At all times the mass in the shell will be dominated by the swept up ISM mass not the integrated wind mass from the star. Thus, we can define the total shell mass as
\beq
\label{shellmass}
M_s = \frac{4}{3}R_s^3 \rho_0,
\eeq
where $\rho_0$ is the average cloud density (pre-wind), $R_s$ is the shell radius, and we assume the shell is roughly spherical. The expansion momentum of the shell is given by
\beq
M_s V_s = (\dot m_w v_w) t_w,
\eeq
where $V_s$ is the shell expansion velocity, $\dot m_w$ is the mass-loss rate, $v_w$ is the wind velocity and $t_w$ is the age of the shell (length of time the wind is driven by the star). This time, $t_w$, can be related to the size and expansion velocity of the shell.  At early times, the expansion follows the self-similar Taylor-Sedov blast wave solution (radiative losses are negligible): $R_s \propto t_w^{2/5}$ \citep{shutextbook}. This corresponds to a shell velocity $V_s = 2/5 (R_s/t_w)$, which gives:  
\beq
M_s V_s = (\dot m_w v_w) \frac{2R_s}{5V_s}.
\eeq
Then the mass-loss rate is:
\beq
\dot m_w =  \frac{10 V_2^2 R_s^2 \rho_0}{3v_w},
\eeq
where the mass of the shell has been replaced using Equation \ref{shellmass}.
The mass-loss rate can be normalized in terms of fiducial values:
\beqa
\dot m_w &=& 5.24 \times 10^{-7 } \left( \frac{V_s}{1~ {\rm km s}^{-1}} \right)^2  \left(  \frac{200~ {\rm km s}^{-1}}{v_w}  \right) \left( \frac{R_s}{1~{\rm pc}} \right) ^2 \nonumber \\
& & \times \left( \frac{n_0}{200~ {\rm cm}^{-3}} \right) \msun {\rm yr}^{-1},
\eeqa
where $n_0$ is the mean particle number density of the cloud $n_0 = \rho_0/m_p$. Here, $1\kms$ is a typical measured shell expansion velocity and $1$ pc is a typical shell radius.


\section{Results}\label{results}

We analyze the simulation outcomes by computing the bulk cloud properties, turbulence statistics, and synthetic observations of the clouds.

\subsection{Bulk Properties}

\begin{figure*}
\vspace{-0.2in}
\plotone{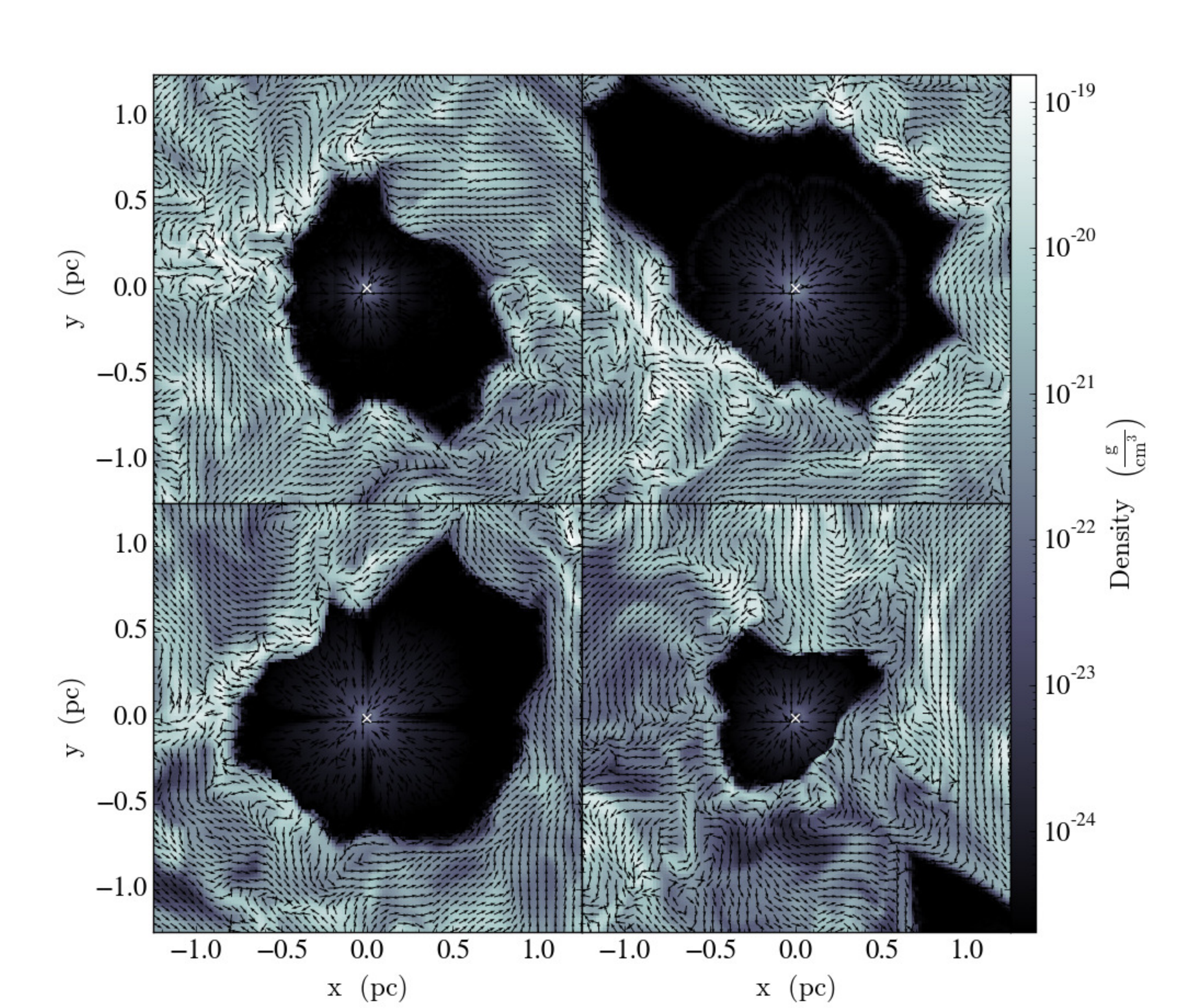}
\vspace{-0.02in}
\caption{ Log density slices through four star locations in run W1\_T2 at 0.1Myr.  The vectors show the direction and relative strength of the magnetic field. The star locations are marked with white crosses.} \label{bfield}
\end{figure*}

Figure \ref{vol_rend} shows volume rendered images of the gas velocity for the models at different times. The regions of high velocity gas are roughly centered on the stellar sources and are co-spatial with the hot $10^3-10^4$ K gas. The shells are more spherical at early times and loose some symmetry as they expand into lower density regions. The magnetic field has a confining and stabilizing effect, which suppresses shell instabilities that might otherwise occur at the interaction boundary between dense and light fluids \citep{krumholz09,arthur11}. Figure \ref{bfield} shows the magnetic field vectors overlaid on slices of the gas density at four star locations. The shells impact the local magnetic field morphology such that the field is mostly parallel to the shell boundaries.  \citet{ridge06b} note a similar configuration for the shell near the IC348 cluster in Perseus, where polarization measurements indicate the magnetic field aligns with the shell boundary.   As gas is swept up by the expanding shell, the bulk magnetic field, which is coupled to the gas, also declines within the evacuated region. Figure \ref{bfield} also illustrates that the shell morphology is a strong function of the local gas density and turbulent distribution. 

A close examination reveals that the shell walls are not completely smooth. A few tendrils of gas on the shell boundaries protrude into the hot, low-density region (e.g., see the right portion of the shell in the top-left panel).  These features increase  with magnetic field strength, leading to noticeable wiggles in the shell surface, and disappear in the weakest field case. We conclude they are likely caused by a magnetic kink instability, which also develops in astrophysical jets \citep[e.g.,][]{lee15}.   We might expect to see Rayleigh-Taylor instability, which frequently occurs at the interfaces between low-density and high-density fluids,  or Kelvin-Helmholz instability, which results from shear between two fluid layers \citep[e.g.,][]{ntormousi11}. However, we do not see any clear evidence of either instability. This may be because the gas is turbulent or because they are suppressed by the magnetic field. 



Runs W3\_T2 and W4\_T2 follow the evolution of individual isolated stars with mass-loss rates of $10^{-9}$ and $10^{-8} $ \msun yr$^{-1}$, respectively. These winds have such a minor sphere of influence that the high-velocity bubble region appears only very faintly in the middle and bottom panels of Figure \ref{vol_rend}. 
As illustrated in Table \ref{bulkres} each of the wind models have very different hot gas filling fractions, which correlate roughly with the total mass-loss rate. 




\begin{deluxetable*}{lccccccc}
\tablecolumns{8}
\renewcommand{\tabcolsep}{0.1cm}
\tablecaption{Bulk Results\tablenotemark{a} \label{bulkres}}
\tablehead{ \colhead{Model} &  
   \colhead{$R_{\rm B_{rms}}$ } &
 \colhead{$R_\mathcal{M}$  } & 
    \colhead{ $f_{V}$} &
    \colhead{$f_{P,i, \sigma_{1D}>2.5 }$ } &
    \colhead{$f_{P,f, \sigma_{1D}>2.5 }$} &
     \colhead{$f_{E,i, \sigma_{1D} > 2.5}$} &
     \colhead{$f_{E,f, \sigma_{1D}>2.5 }$   } }
\startdata
W1\_T1\_L0 & 1.12     & 1.23  & 0.11  & 0.011          &  0.083          &   0.082           &  0.39 \\ 
W1\_T1      & 1.13    & 1.24   & 0.11  &  0.011  &  0.082  & 0.082  & 0.40 \\ 
W0\_T2     &  1.00      & 0.92            & 0.00        &	0.018	& 0.0079	     &	0.037	   &	0.016	\\
W1\_T2     & 1.05    & 1.25    & 0.094     & 0.018 & 0.094 &  0.037     & 0.44  \\
W2\_T2     &  1.00      & 1.00   & 0.016   & 0.018  & 0.018  & 0.037  & 0.063 \\
W2\_T3     & 1.00      & 1.00   & 0.015    &  0.014 & 0.017 & 0.029 & 0.060 \\ 
W2\_T4     & 1.00      &  0.95 &   0.013   & 0.020    & 0.015   & 0.054   & 0.058          \\
W3\_T2     & 1.00       & 1.00  & $1.5\times 10^{-5}$ & 0.018   & 0.018 & 0.037 & 0.037 \\ 
W4\_T2     & 1.00   &1.00      & $2.1\times 10^{-4}$  & 0.018 & 0.018 & 0.037 & 0.037  
\enddata
\tablenotetext{a}{Model name,  the ratio of the rms magnetic field strength at $t_f=0.1$ Myr to $t_i=0$ Myr, the ratio of the rms 3D Mach number at $t_f$ to $t_i$, volume filling fraction of the wind at $t_f$  (defined as the volume of cells for which $T>100$K), momentum fraction of cells with a 1D rms velocity $\ge 2.5 \kms$ at the initial and final times, and the energy fraction in cells with a 1D rms velocity $\ge 2.5 \kms$ at the initial and final times.}
\end{deluxetable*}

Figure \ref{bulk_prop_fig} shows the evolution of the total velocity dispersion, rms magnetic field strength and effective size of the bubble region. Only the highest mass-loss rate case (Run W1\_T1 ) shows appreciable energy injection by the winds, as indicated by the 50\% increase in the gas velocity dispersion. However, after an initial phase of rapid growth, the shells expand more gradually. The shell around the source with the highest mass-loss generally dominates the bulk properties. 

Run W1\_T1 is the only run which exhibits a noticeable change in the rms magnetic field. The magnetic field increases because the field is swept up and compressed by the shell expansion. The diffuse gas in the bubble interiors have negligible magnetic field. Figure \ref{bulk_prop_fig} shows that the evolution of the runs with and without AMR refinement (run W1\_T1 and run W1\_T1\_L0) are nearly identical, indicating that these results are fairly independent of the shell resolution.

In the lowest mass-loss cases, as illustrated by runs W3\_T2 and W4\_T2,  the bubbles rapidly reach pressure equilibrium with the surrounding gas and expansion stalls. Figure \ref{bulk_prop_fig} indicates the time for radiative bubbles to become pressure-confined as predicted by \citet{koo92}, where we modify the prediction to account for turbulent and magnetic pressure (see Appendix \ref{appendix1}). We find the time at which expansion stalls is in good agreement with the predictions for the mass-loss rates. Runs W3\_T2 and W4\_T2 expand slightly longer since their peak wind speed is $\sim600-700\kms$ rather than 200$\kms$, which gives a slightly longer confinement for their mass-loss rates. In the lowest mass-loss case ($\dot m_w=10^{-9}$ \msun yr$^{-1}$) the bubble volume increases and decreases in size depending on fluctuations in the local turbulence and gas density. Between 0.1 Myr and 0.2 Myr the shell becomes so small that it is invisible on the scale of the image in Figure \ref{vol_rend}. 

The \citet{koo92} model predicts that a source with $\dot m_w = 10^{-6}$ \msun yr$^{-1}$ will become pressure confined at 0.22 Myr for our fiducial parameters, which means that all sources with mass-loss rates at or below this value will have stopped expanding by this time.  According to the expansion model, if the bubble remains radiative, at 0.22 Myr the shell radius will be $R_s = 0.63$ pc. For a source with a $\dot m_w = 2\times 10^{-6}$ \msun yr$^{-1}$, the highest mass-loss source in W2, at the time of pressure confinement its size will be $R_s=0.88$ pc. This is comparable to the radii that we find in the simulation, and Figure \ref{bulk_prop_fig} shows the total combined effective shell volume of the sources in W2 is around 1pc at 0.2 Myr.  Since runs W1 and W2 contain multiple sources, the net expansion is dominated by the strongest wind progenitor. As individual bubbles become pressure confined, the net expansion slows. The strongest sources have a mass-loss rate of $1.8 \times 10^{-5}$ and $2.4 \times 10^{-6} \msun$yr$^{-1}$, respectively, which correspond to a expansion time of 0.33Myr and 0.91Myr. The exact time to pressure equilibrium depends on the source location, i.e.,  variation of the ambient density and magnetic field in the star vicinity. 

\subsection{Turbulent Dissipation}

\begin{figure}
\epsscale{1.2}
\plotone{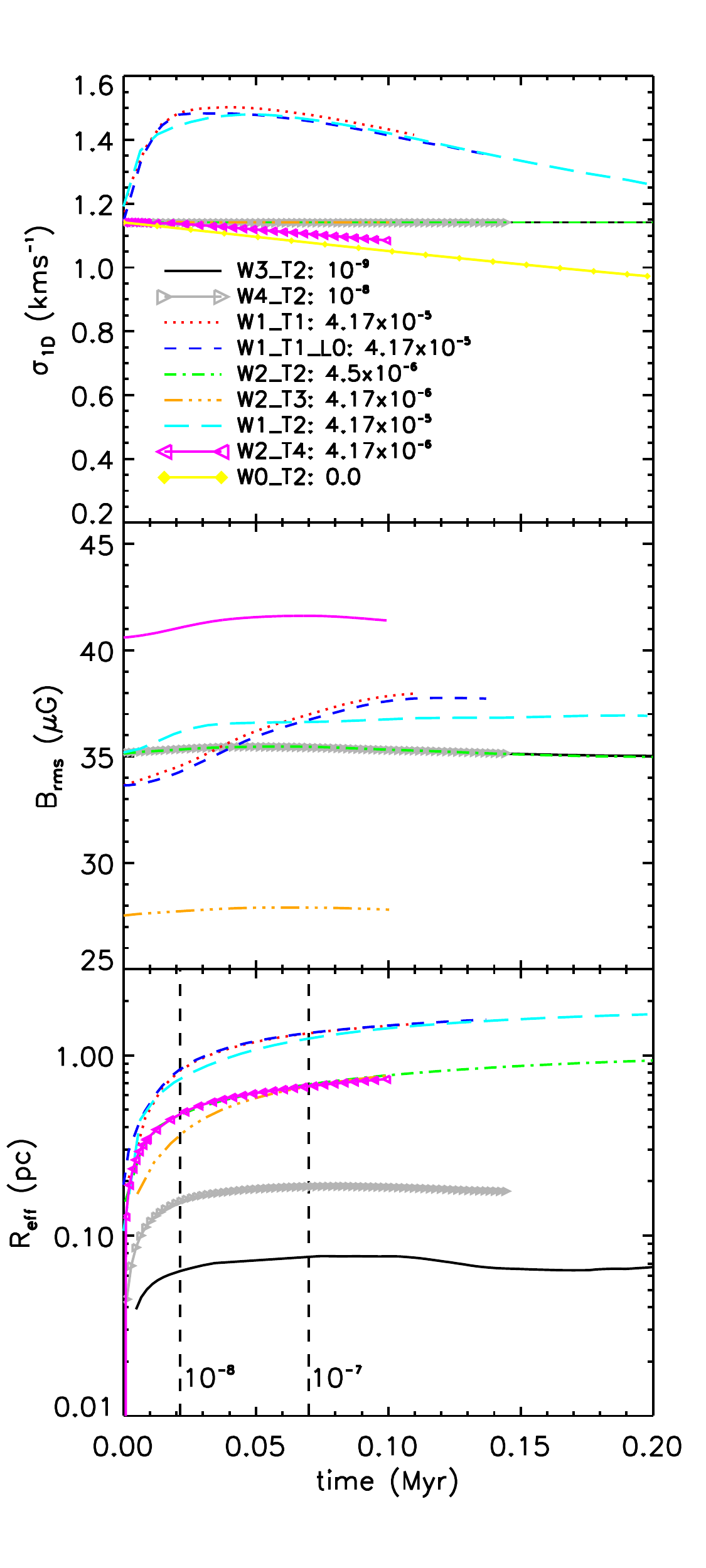}
\vspace{-0.3in}
\caption{Velocity dispersion (top), rms magnetic field (middle) and effective radius ( defined as $R_{\rm eff }=(3V/4 \pi)^{1/3}$ for the total volume enclosed by shells, bottom) as a function of time for the various models, where 
$B_{\rm rms} = \Sigma_i  (B_{x,i}^2 + B_{y,i}^2 + B_{z,i}^2 )^{0.5} / N$  and $R_{\rm eff} = \left[ \Sigma_i \Delta x_{T>1000 {\rm K}}^3 / (4 \pi / 3) \right]^{1/3}$. The legend also displays the total mass-loss rate for each run in $\msun$yr$^{-1}$. The vertical black dashed lines indicate the time that a source with mass-loss rate $\dot M_w=10^{-8}, 10^{-7}\msun$yr$^{-1}$, respectively, is predicted to become pressure confined for our fiducial parameters ($B=13.5 \mu$G, $c_s=0.19 \kms$, $\sigma_{\rm 1D}=1.14\kms$, and $v_w=200\kms$). } \label{bulk_prop_fig}
\end{figure}

We use the global velocity dispersion to estimate the impact of wind driving on the turbulent dissipation.   Visually, the slope of the velocity dispersion (for runs without external driving) in the top panel of Figure \ref{bulk_prop_fig}  indicates the rate of dissipation. We note that the change in the global velocity dispersion for the run with neither  external driving nor winds following the turbulence initialization (W0\_T2) is very similar to those with winds but no external driving (W2\_T4, W1\_T2, and W1\_T1). 
More quantitatively, we define the average turbulent dissipation simply as $\dot E_{turb} = \frac{E_2-E_1}{t_2-t_1} $ for two different times.  This simply reflects the global change in energy and is an indirect measure of the dissipation which occurs on small scales.  Run W0\_T2, which has no winds or turbulent energy injection,  provides an estimate of the expected dissipation rate,  $\dot E_{turb} \simeq 6  \times 10^{33}$ erg s$^{-1}$. The total initial turbulent energy is $E_{turb}=1.47 \times 10^{47}$ erg, which gives a dissipation time of $t_{\rm diss} = E_{turb}/\dot E_{turb} \sim 0.8 $ Myr. If the dissipation time is $t_{\rm diss} = \eta t_{\rm ff}$ \citep{mckee89,maclow99}, then $\eta \simeq 0.5$,  which is within the expected range. Here, the free-fall time is given by $t_{\rm ff} = \sqrt(3 \pi /(32 G \rho))=$1.47 Myr. 

The model W1 velocity dispersion is initially enhanced by the winds, but begins to show strong dissipation after 0.05 Myr. Using the times between 0.1 Myr and 0.15 Myr, we estimate a dissipation rate of $\dot E_{\rm turb} = 1.68 \times 10^{34}$ erg s$^{-1}$, which is twice as high as the non-wind case. If the dissipation rate scales with the rms velocity, $\dot E \propto \sigma ^3$  \citep{maclow99}, the dissipation rate in run W1 at 0.1Myr should indeed be $(3\sigma_{W1}/\sigma_{W2})^3=(2.5\times 10^5/2.0 \times 10^5)^3 \simeq 2$ times larger than run W0\_T2 at the same time.

The energy injection rate from all winds is $\dot E_w = \frac{1}{2} \Sigma_i \dot m_{w,i} v_{w,i}^2$. Although the wind velocity varies as a function of mass, models W1 and W2 have maximum wind velocities of 200 $\kms$, such that the wind speed is effectively 200 $\kms$ for all stars. Models W1 and W2 have an energy injection rate of $5.6 \times 10^{34}$ erg s$^{-1}$ and $5.2 \times 10^{35}$ erg s$^{-1}$, respectively (see Table \ref{simprop}). Both values exceed the global energy dissipation in the non-wind simulation, which initially might imply that the wind input should be sufficient to maintain the global turbulent energy.  However, since the wind expansion slows as the shells become pressure confined, the high wind velocities and energy input do not directly translate into turbulent motion. 

In Run W2\_T4 the rms velocity goes up by $\sim 0.1$ percent initially, but turbulent dissipation quickly dominates.  From 0.02-0.1 Myr, the turbulent dissipation is $5.2 \times 10^{33}$ erg/s. This is about 15\% smaller than the fiducial value, however more strongly magnetized turbulence has a smaller dissipation rate \citep[e.g.][]{stone98,maclow99}, and this value is consistent with the prior numerical studies of turbulent dissipation. Although the winds do inject energy locally, these results suggest that they are not sufficient to offset the turbulence dissipation and cannot maintain the initial cloud turbulence.

 

\subsection{Turbulent Statistics}

In order to assess the impact of the winds on the cloud properties and distribution, we compute several common turbulent statistics: the Fourier velocity, density, and momentum power spectra and the density, velocity and momentum probability distribution functions (PDF). The choice of these statistics is partially motivated by prior work indicating their utility as diagnostics for various turbulent properties. For example, \citet{swift08}  found that bipolar protostellar outflows can generate a feature in the  power spectrum of the CO integrated intensity. Similarly, \citet{carroll09,carroll10} demonstrated that outflows may impact the velocity power spectrum slope. \citet{federrath10} showed that the mode of turbulent forcing, whether compressive or solenoidal (stirring) also impacts the gas density PDF, where compressive forcing produces a broader range of densities.  Here, we examine whether the wind shells produce additional compressive forcing. 

\begin{deluxetable}{lcccc}
\tablecolumns{5}
\tablecaption{Spectral Slopes\tablenotemark{a} \label{slopes}}
\tablehead{ \colhead{Model} &  
   \colhead{$t$(Myr) } &
 \colhead{$\alpha_\rho$  } & 
    \colhead{ $\alpha_{M}$} &
     \colhead{$\alpha_{v}$} }
\startdata

W2\_T3   & 0.0 & -1.39 & -0.58   &  -1.7\\ 
W2\_T3   & 0.1 & -1.38  & -0.55  & -2.5  \\ 

W1\_T1    & 0.0  & -1.38   &  -0.50    & -1.7 \\
W2\_T2    & 0.0 & -1.33  & -0.56    &  -1.7 \\   
W2\_T2    & 0.1 & -1.30 & -0.50   &  -2.6\\ 

W1\_T2   & 0.1 &  -1.43  & -0.45    &  -2.4   \\  
W1\_T1    & 0.1 & -1.57  & -0.39    &  -2.4\\

W2\_T4  & 0.0 &  -1.39    & -0.39 &  -1.7  \\ 
W2\_T4    & 0.1 &  -1.41   & -0.39  &  -2.7    
\enddata
\tablenotetext{a}{Model name, time, slope of the Fourier spectrum of logarithmic density, slope of the Fourier spectrum of momentum, and slope of the Fourier spectrum of velocity. The analysis employs a uniform 512$^{3}$ grid,  which is interpolated from the AMR data. The fits are performed over the range $k=4.3-16.7$. The $1-\sigma$ uncertainty in the $\alpha_\rho$, $\alpha_M$, and $\alpha_v$ fits are approximately $\pm 0.02$, $\pm 0.05$, and $\pm 0.1$, respectively.  We list the runs in the table in order of increasing magnetic field (see Figure \ref{bulk_prop_fig}).}
\end{deluxetable}

\subsubsection{Probability Distribution Functions}

\begin{figure*}
\epsscale{1.1}
\plotone{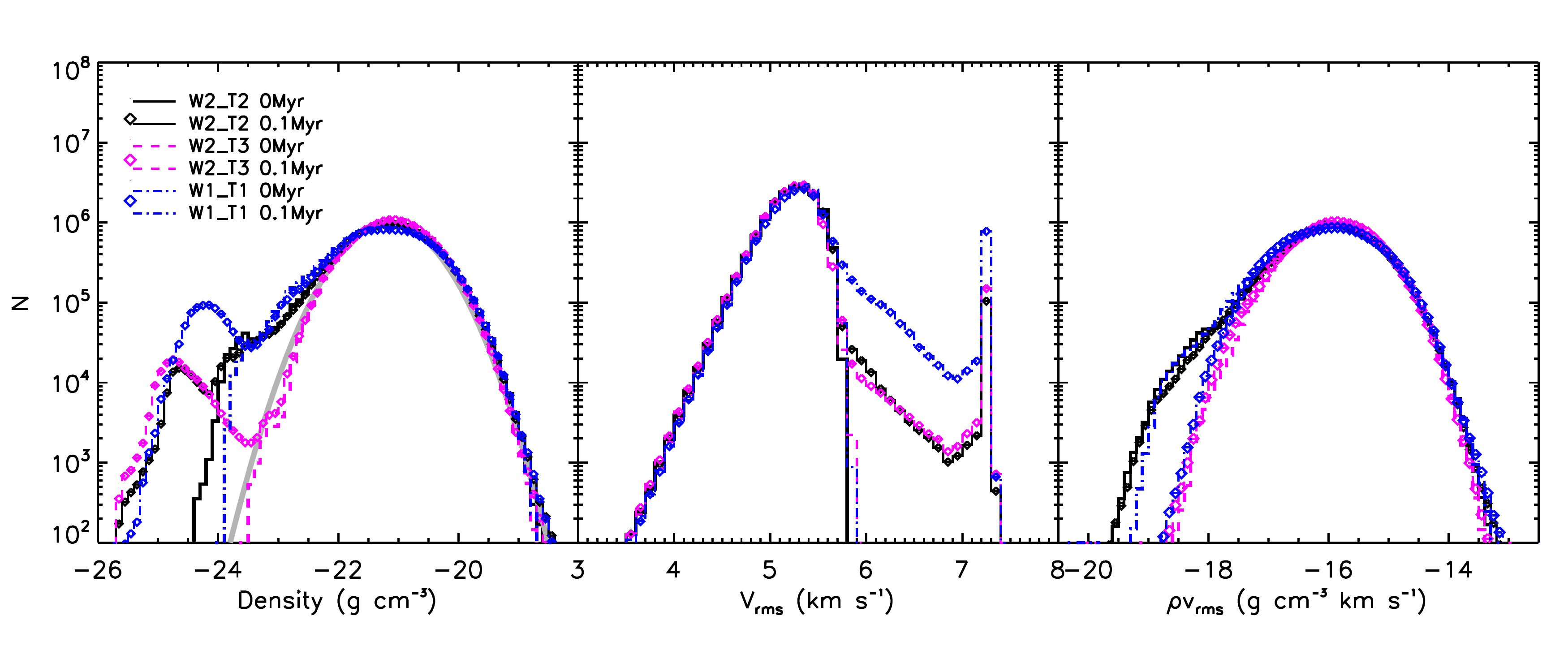}
\vspace{-0.1in}
\caption{Density (left), velocity (middle) and momentum (right) PDFs for different models at times prior to wind launching (0 Myr, lines) and at 0.1 Myr (lines and symbols). A lognormal function in grey is shown in the left plot: $p(ln \rho) = 9.7 \times 10^5 exp \left[ -1.3 (ln \rho - 21.2)^2 \right]$.} \label{pdfs}
\end{figure*}

The density PDF of gas in driven turbulence simulations is generally very close to a log-normal distribution \citep{maclow99,padoan04, kritsuk07,federrath10}:
\beq
p({\xi})d \xi = \frac{1}{\sqrt{2 \pi \sigma^2}} exp\left[ -\frac{(\xi - \bar{ \xi})^2} {2\sigma^2}  \right] d\xi,
\eeq
where $\xi= {\rm ln}(\rho/\bar{\rho})$, $\sigma$ is the width, and normalization ensures that $\int p(\xi) d\xi = 1$. The top panel of Figure \ref{pdfs} shows the density PDFs with and without winds. The non-wind turbulent density PDFs resemble a log-normal distribution as expected. However, once wind launching commences gas is evacuated from a fraction of the domain, leaving low-density voids. This manifests as a second peak at densities of $\rho \sim 10^{-24}-10^{-25}$ g cm$^{-3}$. However, gas redistribution appears to have little impact at high-densities, i.e., the mass that piles up in the shells does not modify the global density distribution on large scales.

The velocity PDF,  shown in the middle panel of Figure \ref{pdfs}, exhibits the most difference in the case with winds. It develops a high-velocity tail that extends to several hundred $\kms$. The peak at 200 $\kms$ corresponds to the wind launching velocity maximum set in the simulations. As expected, the runs with the highest total mass-loss generate the largest amount of fast moving, low density gas. Otherwise, the distributions remain very similar to the initial turbulent PDFs. 

The third panel of Figure \ref{pdfs} shows the momentum PDF.  All momenta PDFs appear to be approximately log-normal, with similar widths. Some asymmetry in the density and velocity  distribution appears as a result of the random driving.  The PDF of the gas with and without winds is nearly indistinguishable. Essentially, the wind shells are shocks, which have very similar momenta to the initial turbulence.  The shell expansion velocities are $\sim 1-3~ \kms$, which bracket the turbulent velocity peak at 1.2 $\kms$. This accounts for the minimal change in the bulk velocity and momentum distributions. Likewise, the PDFs for the strong magnetic field case are not statistically different than the weaker field case.

If the winds are preferentially adding compressive modes, then we might expect the density and momenta PDFs to significantly broaden \citep[e.g.,][]{federrath08,federrath10}. However, aside from the appearance of a second peak in the density PDF, this does not occur to any significant degree. Quantitatively, the standard deviation of the momentum distribution increases by 7\% and 0.8\%, respectively, for runs W1\_T1 and W2\_T2.  The outputs with winds are otherwise similar to the pre-wind turbulence, even in the case with the highest mass-loss rates, in which the winds have a volume filling fraction $>10$\%. The similarity of the pre- and post-wind runs could also be because the winds do not create non-local effects (see \S\ref{driving}).

\subsubsection{Power Spectra}

We compute the Fourier power spectrum of the density, momentum and energy distribution using \cite[e.g.][]{federrath13}:
\beq
P (q, k) =  \avg{ \hat{q} \cdot{ \hat{q}}^* 4 \pi k^2 }_k ,
\eeq
where $\hat{q}$ is the Fourier transform of a three-dimensional quantity $q(l)$:
\beq
\hat{q}({\bf k}) = \frac{1}{(2\pi L)^{3/2}} \int q(l) e^{i {\bf k} \cdot {\bf l}} d^3l
\eeq
for $l$ in the range $[0,L]$ with the maximum spatial scale L corresponding to the smallest wavenumber $k=2\pi/L$.

Although the effective grid resolution  including AMR is $1024^3$, the basegrid resolution (minimum resolution without AMR) is 256$^3$. This means the initial inertial range is $k \sim 5-20$,  since there is little AMR refinement at early times. The inertial range is limited at small $k$ because we inject turbulence over the wave numbers $k=1-2$.
Once wind driving begins, the wind shells and strong shocks will be refined to higher resolution, increasing the effective inertial range. We perform the analysis using flattened 512$^3$ grids for all the variables. A comparison of 256$^3$ and 512$^3$ resolution is presented in Appendix \ref{appendix2}. 
We summarize the best-fit slopes for each of the Fourier spectra in Table \ref{slopes}.

\begin{figure}
\vspace{-0.25in}
\epsscale{1.17}
\plotone{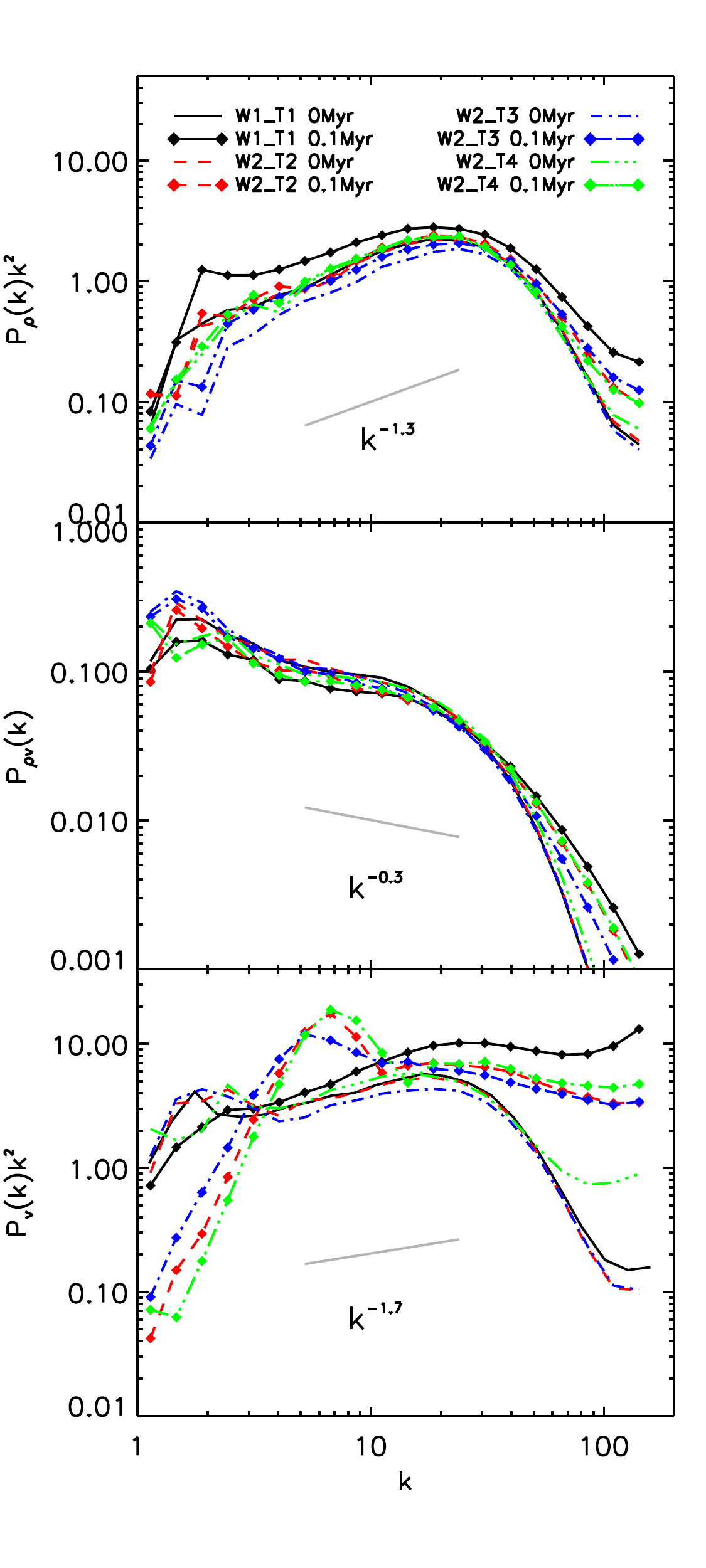}
\vspace{-0.25in}
\caption{Log density (top), momentum (middle) and velocity (bottom) Fourier spectra computed using a 512$^3$ grid resolution. The grey line denotes the approximate inertial range and indicates the typical best fit slope. The density and energy spectra are compensated by $k^2$. The input quantities are normalized to the average density, rms momentum, and rms velocity, respectively.} \label{power}
\end{figure}

The top panel of Figure \ref{power} shows the Fourier spectrum of logarithmic density. The spectral slopes are comparable to $S(k) \propto k^{-1.3}$, with the winds causing a slight steepening in the slope in the cases with strong winds (W1). The fitted slopes are all consistent with the solenoidal driving case in \citet{federrath10}, for which $\alpha_{\rho} \propto k^{-1.5}$. Compressively driven turbulence exhibits a significantly steeper density spectrum: $\alpha_{\rho} \propto k^{-2.3}$. Since the W2 snapshots with winds are not significantly statistically different over the inertial range, this suggests the turbulence is still largely solenoidal. The slope of Run W1 does steepen slightly, which may indicate moderately enhanced compression.  The winds impact the shape of the spectrum over the dissipation range (i.e., $k=40-100$), which becomes slightly flatter in the cases with winds. This does not appear to be a product of the increased resolution in the wind runs since the same difference appears for power spectra computed at a lower grid resolution (see Appendix \ref{appendix2}) and for the run W1\_T1\_L0.

The Fourier spectra of the momentum ($q=\rho {\bf v}$) are shown in the middle panel of Figure \ref{power}. These spectra are likewise very similar over the inertial range for the wind and non-wind cases.  The only difference attributable to winds manifests over the dissipation range at larger $k$ values. The outputs with winds are again slightly systematically higher (more momentum on small scales) than the same model without winds.

The velocity power spectra (bottom panel of Figure \ref{power}) is the only statistic to show a strong departure from the non-wind case. A peak occurs at low k values, $k\sim4-10$, which is indicative of the bubble size, with the strong wind case peaking at the smallest k ($l \sim L/4$).  This creates a pronounced steeping of the slope, $\alpha_v \simeq -2.4$,  for a power-law fit over the inertial range. As in the other cases, the winds impact the dissipation scale by adding a little extra power on small scales.

 
The power-law slopes are nearly independent of the average magnetic field strength. Only the Fourier momentum spectral slope exhibits a statistically significant change: the slope flattens with increasing magnetization. However, the wind feature in the velocity spectrum noticeably shifts to smaller $k$ with decreasing field strength. This is indicative of the role of magnetic pressure confinement in setting the bubble size. Consequently, stronger field strengths correlate to slightly steeper slopes.
 
 All the velocity slopes are flatter than the solenoidal and compressive driving slopes found by \citet{federrath10}, $\alpha_v \simeq $ -1.8 to -2.0 for non-magnetized turbulence. In general, the velocity power spectral slope is expected to fall between $-\frac{5}{3}$ \citep{kolmogorov41} and $-2$ \citep{burgers48}.  The difference may be partially due to lower resolution, which is discussed in Appendix \ref{appendix2}.

\subsection{Synthetic CO Distributions}

\subsubsection{CO Modeling Method}

We post-process the simulations with the radiative transfer code {\sc radmc3d}\footnote{http://www.ita.uni-heidelberg.de/~dullemond/software/radmc-3d/} to obtain the emission in the $^{12}$CO(1-0) molecular line. We compute the emission using the Large Velocity Gradient (LVG) approximation \citep{shetty11}, which calculates the rotational level populations by solving the equations for local radiative statistical equilibrium. We perform the radiative transfer using the densities and velocities on a $256^3$ grid, where we define $n_{\rm H_2} = \rho/(2.8 m_p)$ and adopt a CO abundance of [$^{12}$CO/H$_2$] =$10^{-4}$ \citep{frerking82}.  Gas above 800 K or with $n_{\rm H_2} < 10$ cm$^{-3}$, i.e. effectively all gas in the rarefied bubble regions, is assumed to have a CO abundance of 0. Gas with densities $n_{\rm H_2} > 2 \times 10^4$ cm$^{-3}$ experiences freeze-out onto dust grains \citep{tafalla04a}, which significantly reduces the gas phase CO. We set the abundances to zero in these high density regions.  We include turbulent line broadening on scales at and below the grid resolution by adding a constant micro-turbulence of 0.25$\kms$, which is determined by the observed linewidth-size relation (see \S\ref{global_param}). The spectral resolution of the output cube is $\Delta v = 0.156~ \kms$.

To model an observation of Perseus, we assume the simulated clouds are located at a distance of 250 pc and convolve the emission with an 46'' beam, which is the resolution of the COMPLETE $^{12}$CO (1-0) survey of Perseus \cite[e.g.,][]{ridge06}.

Since the morphology of the observed bubbles suggests that Perseus is flatter along the line of sight than its projected length (e.g., A11 conclude $l_{\rm los} \lesssim 0.5$pc), we also calculate a spectral cube for a thin cloud, which is a $L_z=0.8$ pc wide subset of the whole domain (Run W1\_T1\_S1). We choose the subdomain such that some of the bubbles span the entire cloud depth and some sight lines contain no molecular emission. Indeed, some of the larger bubbles observed in Perseus ``break out" of the cloud and do not enclose molecular gas. Similar cloud-spanning bubbles have also been identified in other clouds \citep{beaumont10,deharveng10,nakamura12}.

\subsubsection{Synthetic CO Intensity Maps}

Figure \ref{cotb} shows maps of the integrated brightness temperature. Some of the wind shells appear clearly in the CO emission, although not all resemble rings due to superposition or interaction between the winds. The CO line is strongly excited in the shells because the gas is both denser and warmer than the mean cloud density ($n_{\rm H_2}\simeq 400$ cm$^{-3}$) and temperature ($T=10$ K).  At 0.1 Myr the shell sizes are typically less than 1.5 pc with a strong dependence on the source mass-loss rate and some dependence on the ambient density near the source. The projected radii of the CO shell emission in Perseus are $\sim$0.2-2.8 pc, which is comparable to the sizes of the shells modeled here. Given an expansion rate of 0.1 pc per 0.1 Myr  for the strongest source in model W2 (see Figure \ref{bulk_prop_fig}), the radius of the largest shell might exceed 2 pc following another Myr of evolution.

Although the shells do not resemble circular rings, they are distinctly rounded structures.  The shells appear more clearly in the case where the bubbles span the width of the cloud.  Like observed shells, the simulated structures have variable brightness, i.e., they are clumpy. Here, this is caused by density variance in the turbulent gas, such that different parts of the shell interact with denser ambient material. If the ambient gas is lower density then the hot gas expands more freely, leading to additional asymmetry. These characteristics agree with the observations of Perseus, for which not all shells are complete circles or are easily identified in the integrated emission map; some must be identified on the basis of their coherent velocity structure in the spectral cube. The simulated sources with weaker winds, $\dot m_w \lesssim 10^{-7} \msun$yr$^{-1}$, also produce much weaker shell features, which are not necessarily visable in the CO emission.

Wind model W2, which has sources with similar mass-loss rates as those estimated in Perseus, also exhibits shell structure with similar CO intensity. As reported in A11, the identified Perseus shells have average CO integrated intensities of $\sim 0.6-2.4$ K $\kms$ (see their Figure 4). 
 The shells in the simulated W2 maps, which have been convolved to similar spatial and velocity resolution as the observations, have CO integrated intensities in the range of 0.7 to 2.1 K$\kms$, similar to that observed in Perseus.
Without winds, the CO integrated intensities do not exceed 0.7 K $\kms$. In the strong wind case, W1,  shell emission ranges from $0.7-7.0$ K $\kms$, which exceeds the shell emission in Perseus. This supports the estimate by A11 that sources in Perseus have mass-loss rates from $10^{-7}\msun$yr$^{-1}$ to a few times $10^{-6}\msun$yr$^{-1}$, which is consistent with run W2.  For wind velocities of $200 \kms$, model W2 has an input momentum of $\dot P = \dot M v_w =  9 \times 10^{-4} \msun \kms$yr$^{-1}$.  Although we can't directly confirm that all observed momentum is generated by line-driven winds (see \S\ref{altscenario}), we have high confidence in the magnitude of the momentum input needed to produce the observed features.
The stronger sources in model W1 produce features that are too bright, so it is unlikely that any of the Perseus sources have mass-loss rates exceeding $5 \times 10^{-6}\msun$yr$^{-1}$, in good agreement with the observed estimates.


\subsubsection{Hot Gas Distribution}

\begin{figure*}
\plotone{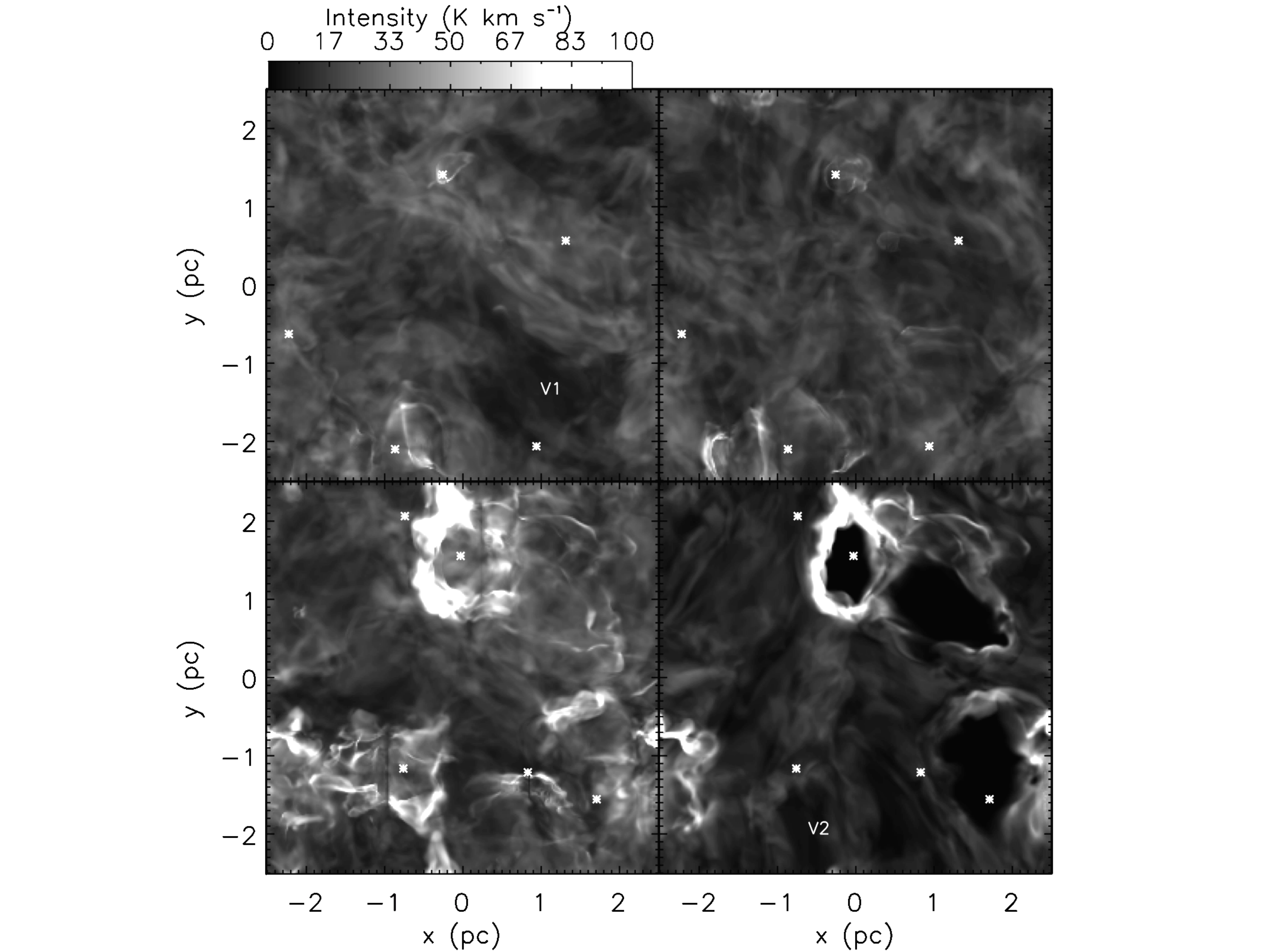}
\caption{Integrated $^{12}$CO(1-0) brightness temperature for Run W2\_T2 (top left), Run W2\_T3 (top right), Run W1\_T1 (bottom left) and a 0.8pc slice of Run W1\_T1 (W\_1T1\_S1, bottom right) at 0.1 Myr.  The projected locations of the stars are marked with asterisks.  V1 and V2 mark two voids discussed in the text.}\label{cotb}
\end{figure*}

\begin{figure*}
\plotone{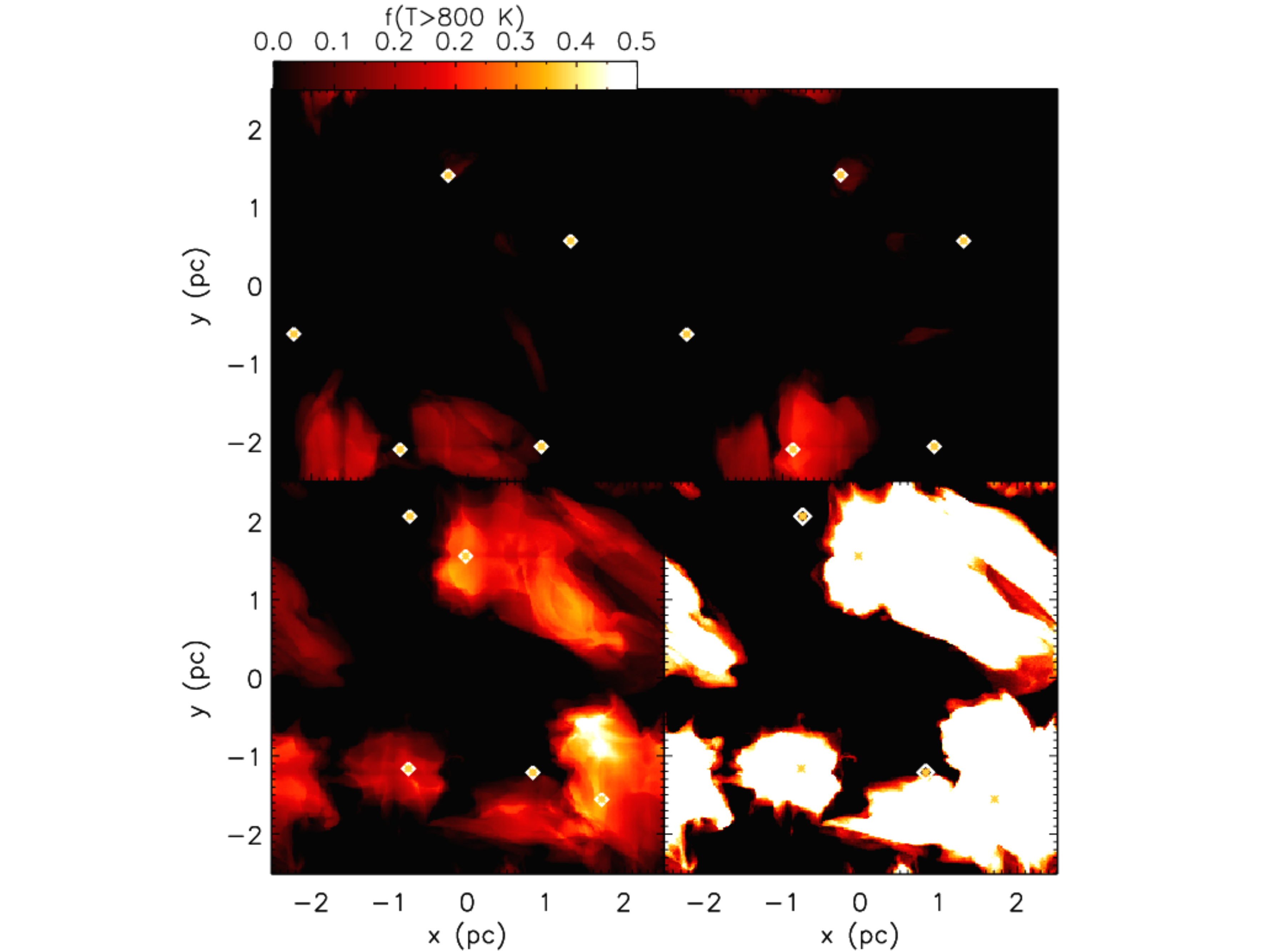}
\caption{Integrated temperature mask including all gas with $T>800$ K for Run W2\_T2 (top left), Run W2\_T3 (top right), Run W1\_T1 (bottom left) and a 0.8pc slice of Run W1\_T1 (W\_1T1\_S1, bottom right) at 0.1 Myr.  The colorscale indicates the fraction of the cloud along each sightline with temperature above 800 K. The projected locations of the stars are marked with asterisks.  The image is saturated where the shells have broken-out of the cloud.}\label{tmask}
\end{figure*}

As discussed by A11, some of the identified shells are ambiguous. Either their signature in the velocity channels is less clear or there is no obvious co-spatial wind progenitor. Turbulence, which naturally produces spatial fluctuaions over large scales, provides one alternative explanation for these features. Statistical fluctuations in the density and velocity distribution may produce voids, which could masquerade as wind features. To test this possibility, we examine the distribution of hot gas. Figure \ref{tmask} shows an integrated mask selecting gas with temperatures greater than 800 K. Since gas is only heated through wind interactions, gas temperature serves as more direct tracer of shells than gas density. Comparison of Figures \ref{cotb} and \ref{tmask} show that most sources are indeed associated with regions of hot gas. However, in many cases the hot gas is able to leak out of the immediate vicinity of the source and create significantly extended and asymmetric bubbles. The hot gas quickly expands in lower density regions, so the bubble shapes are in some sense indicative of the pre-wind turbulent structure. The asymmetric expansion of the bubbles leads to the appearance of distinct shells without any central source (e.g., top-right feature in the images showing run W1\_T1). Because the domain is periodic some hot gas also appears on the opposite side of the domain, which highlights the large extent of some of the bubbles. 

Comparing the hot gas distribution in Figure \ref{tmask} with the apparent voids (dark regions) in Figure \ref{cotb} indicates that many but not all apparent voids are associated with hot wind gas. For example,  V1 and V2 indicate two evacuated regions that do not have corresponding hot gas in Figure \ref{tmask}. This suggests that some small fraction of the shells may be caused by turbulent stochasticity rather than winds.
Unfortunately, the temperature of low-density material is difficult to probe  observationally  when there is little or no CO emission present. Furthermore, even in cases where atomic emission is detected, the gas may be warm because it has been dissociated by the background stellar FUV field rather than by heated by local sources.


\subsubsection{CO Distribution}

\begin{figure}
\epsscale{1.2}
\plotone{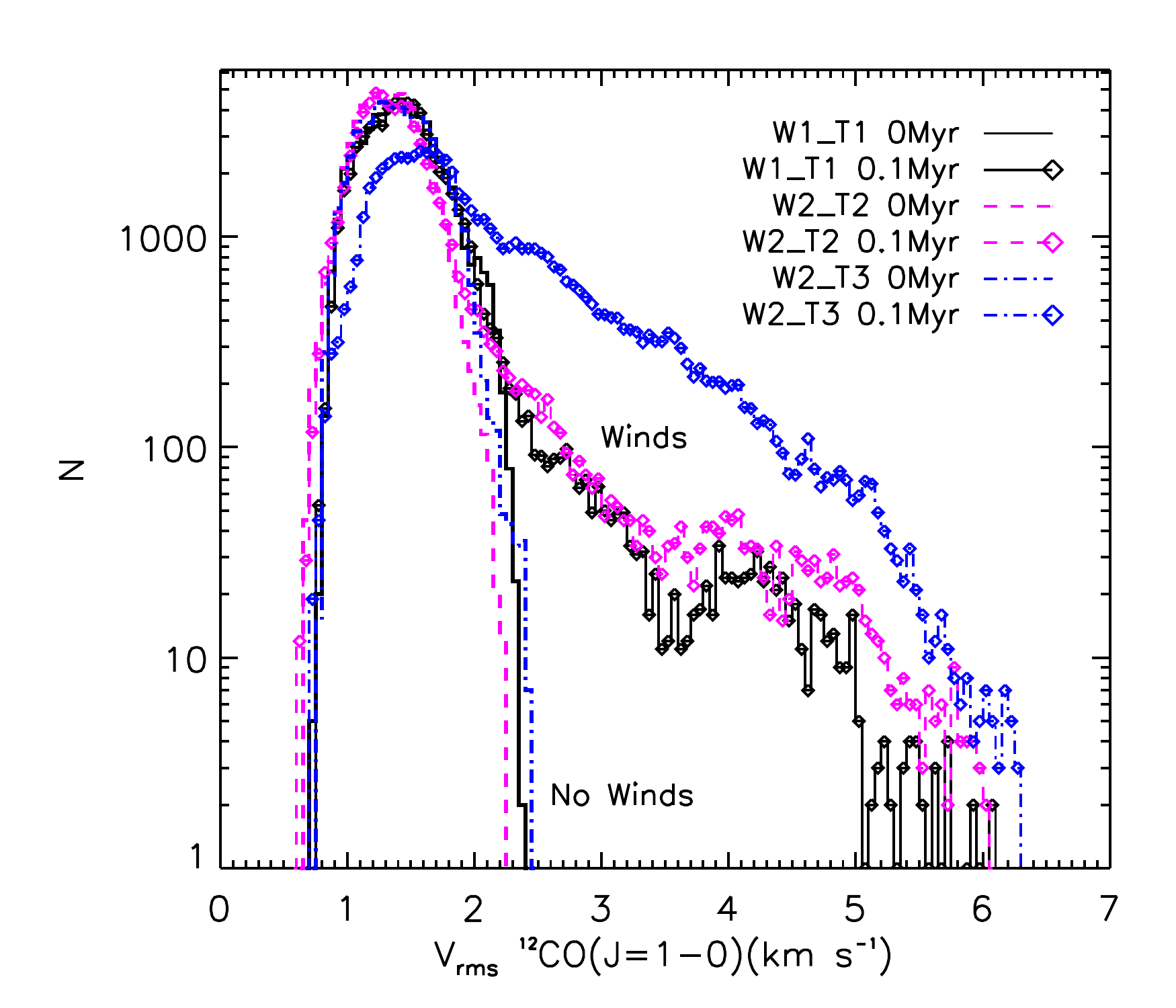}
\caption{Histograms of the intensity-weighted rms velocity, $v_{\rm rms}$, for all sight lines through the cube. The spectral cubes have been convolved with a beam of 46'' at d=250pc. The symbols indicate the snapshots with winds. }\label{codist}
\end{figure}

Figure \ref{codist} shows the rms velocity distribution of the smoothed CO emission for snapshots with and without winds. The rms velocity, $v_{\rm rms}$, is defined as the second moment of the intensity distribution along a given sightline:
\beq
v_{\rm rms} = \left [ \sum_k T_{B,k} (v_k - v_{\rm cent})^2/ \sum_k T_{B,k} \right]^{0.5},
\eeq
where $v_{\rm cent}$ is the centroid velocity or first velocity moment:
\beq
v_{\rm cent} = \sum (T_{B,k}v_k ) / \sum T_{B,k}.
\eeq
  As illustrated by Figure \ref{codist}, winds create a distinctive tail in the velocity distribution.  Although most of the tail is only a few $\kms$, it extends the distribution significantly above the median cloud velocity. This range of velocities, $3-5~ \kms$,  is also consistent with the measured expansion velocities of observed shells (e.g. A11). We also note that the shape of the velocity feature is similar to that of the observed velocity distribution in Perseus \citep{beaumont13}, where it was conjectured to arise from stellar feedback. Table \ref{coprop} gives the mean, median, and standard deviation of the $v_{\rm rms}$ distribution. The cases with winds have systematically higher mean and median  $v_{\rm rms}$ values and have standard deviations as much as three times higher. We note the highest velocity material generated by the winds occurs in hot (ionized or atomic) gas and thus is assumed to have zero CO abundance by construction.

Assuming that the integrated intensity is a proxy for mass, we can estimate the fraction of momentum and energy in high-velocity gas.
 Prior to wind launching less than 0.1\% of the momentum or energy is in emitting gas with a line-of-sight velocity  $v_{\rm rms}>1 \kms$. However, Table \ref{coprop} shows that winds introduce higher velocity material and a significant fraction of both the momentum and energy falls above this threshold. The exact amount of high-velocity gas is very sensitive to the total mass-loss rate, with up to 50\% of the momentum being in winds for the highest mass-loss rates. For a mass-loss rate similar to that estimated for Perseus,  10\% of the momentum is above this threshold and 20-30\% of the energy. The cloud slice (W1\_T1\_S1) and full cloud have a similar fraction of energy and momentum above the threshold.

\subsubsection{CO Turbulent Statistics}

Occasionally, analysis of the integrated CO emission or column density can provide insight into the underlying turbulence or star formation activity  \citep{swift08,carroll10, burkhart10}. NGC1333, which exhibits copious outflow activity shows a spectral break around $k\simeq20$ in $^{13}$CO(1-0). No break is apparent in the $^{12}$CO emission, although this may be attributed to the higher optical depth of $^{12}$CO.

Following \citet{swift08} we calculate the power spectra of the integrated intensity for $^{12}$CO and $^{13}$CO. This entails performing a Fourier transform of the 2D intensity map, squaring the result, and then averaging over annular rings in $k$ space. We subdivide the integrated emission into red-shifted and blue-shifted gas, which spans the range of shell velocities.  For the simulated clouds, there is no net global velocity, so the red- and blue-shifted ranges are simply defined symmetrically about 0 $\kms$.

\begin{figure}
\epsscale{1.24}
\plotone{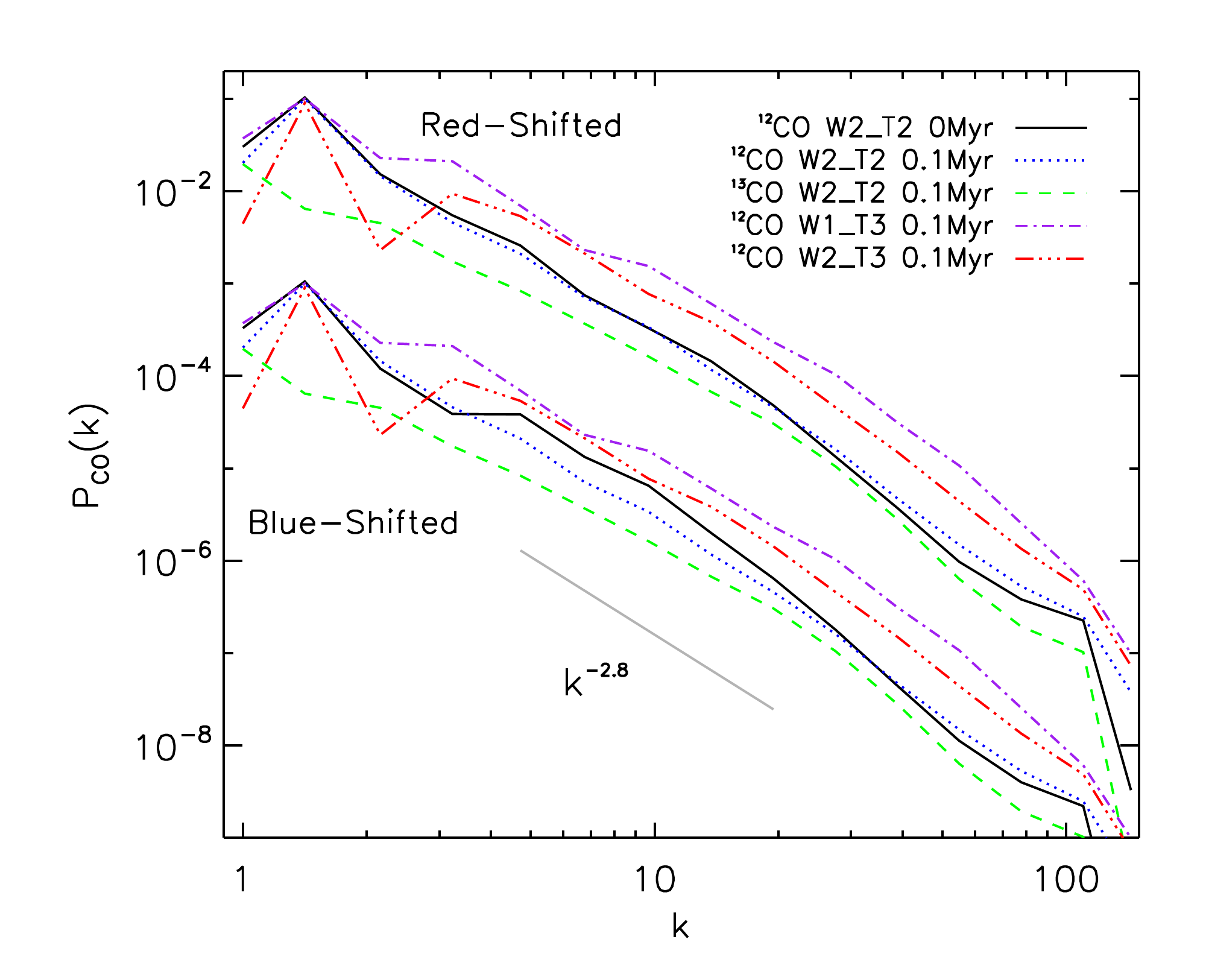}
\caption{ Power spectrum of the integrated CO intensity for gas with velocities $-4.5 \leq v \leq 1.0 \kms$ (blue-shifted) and $1.0 \leq v \leq 4.5 \kms$ (red-shifted). The blue-shifted spectrums are offset by a factor of 50 for clarity. Power spectra are calculated for both $^{12}$CO(1-0) and $^{13}$CO(1-0) (green-dashed line) for W2\_T2 at 0.1 Myr.   }\label{copower}
\end{figure}

Figure \ref{copower} shows the power spectra of this emission for various wind models. We find the red- and blue-shifted emission have similar slopes and none of the models with winds show any clear feature in the spectrum that indicates a preferred energy injection scale. The $^{12}$CO line wings of the model with only turbulence ($t=0$Myr) have slopes of $-2.7 \pm 0.1$ and $-2.8 \pm 0.1$, while W2\_T2 and W2\_T3 have slopes ranging from $-2.5$ to $-2.7$. These slopes are slightly shallower, but consistent within $1\sigma$ error.  The strongest wind case, W1, has $^{12}$CO line wing slopes of $-2.3 \pm 0.2$ and $-2.54 \pm 0.05$, which is marginally flatter than the weaker wind runs and could indicate a trend with increasing wind activity.

The  $^{13}$CO line wings each have a slope of $-2.35 \pm 0.02$. This result is consistent with the expectation that optically thick lines saturate at a slope of $-3$, while optically thiner lines, like $^{13}$CO have flatter slopes \citep{lazarian04,burkhart13}.  Additional analysis exploring different velocity ranges, the full integrated intensity cube, and the gas column density likewise reveal no clear break in the intensity spectrum. The corresponding slopes for different velocity ranges tend to vary by an additional $\pm 0.1$ with higher velocity channels with less average emission producing flatter slopes.

While the lack of a feedback signature may be surprising, breaks in observed spectra are rare and seem to appear only in regions with very vigorous feedback, such as NGC1333. No observational data to date has revealed any such feature attributable to feedback from winds. 

\begin{deluxetable*}{lcccccc}
\tablecolumns{7}
\tablecaption{Synthetic $^{12}$CO Emission Properties \label{coprop}}
\tablehead{ \colhead{Model\tablenotemark{a}} &  
   \colhead{ $t$(Myr) } &
  \colhead{ $\bar v$($\kms$)  } &
 \colhead{ $v_{\rm med}$($\kms$) } &
    \colhead{ $\sigma_v$($\kms$)} &
 \colhead{ $P_{|v_{\rm rms}|>1}/P_{\rm tot}$} &
   \colhead{ $E_{|v_{\rm rms}|>1}/E_{\rm tot}$  }}
\startdata
W1\_T1         &   0       &  1.41     &1.40   & 0.29 & 0.0    & 0.0   \\
W1\_T1         &   0.1    &  2.04     &  1.77  & 0.86 & 0.51 & 0.66  \\
W1\_T1\_S1 &    0       &   0.73    & 0.69   & 0.26  & 0.0  & 0.0   \\
W1\_T1\_S1 &    0.1    &  1.02    &  0.82  & 0.73 & 0.29  & 0.50  \\
W1\_T2         &   0.1    &   2.06    &  1.67 &  0.97 & 0.63 & 0.77   \\
W2\_T2         & 0         & 1.43     & 1.41   & 0.29 & 0.0   & 0.0  \\ 
W2\_T2         &  0.1     &  1.49    & 1.43   & 0.44 & 0.10 & 0.21 \\ 
W2\_T3         &  0        & 1.33    & 1.33   & 0.25 & 0.0   & 0.0   \\
W2\_T3         &  0.1      & 1.45   & 1.35   & 0.53 & 0.14 & 0.30  \\
W2\_T4         &  0         &  1.43   & 1.35  & 0.30   & 0.0   & 0.0   \\
W2\_T4         &  0.1      &  1.46   & 1.39  & 0.48   & 0.11  & 0.25       
\enddata
\tablenotetext{a}{Model name, mean rms spectral velocity, median rms spectral velocity, standard deviation of the rms spectral velocity, fraction of momentum in pixels with los rms velocities $|v_{\rm rms}|>1 \kms$, and  fraction of energy in pixels with los rms velocities $|v_{\rm rms}|>1 \kms$. We assume the CO intensity is a proxy for the underlying gas mass  i.e., we assume the emission is optically thin, which is the assumption that underlies the CO X-factor.} 
\end{deluxetable*}
\vspace{0.75in}




\section{Discussion}\label{discussion}

\subsection{Wind Signatures}

Stellar winds create a significant amount of high-velocity gas that leaves a clear signature in the density and velocity PDF, however most of this gas is quite rarefied and is confined to the central regions of the shells. This material, given its density and temperature, is not in molecules and would not show up in CO emission (except as a void). It might be possible to detect these regions in radio or X-rays, however.  The cold molecular material entrained and accelerated by the expanding shells does produce a clear signature, one that is consistent with the characteristics of the observed CO emission in Perseus.

There are other clear differences between the bulk properties of the CO emission and the underlying densities and velocities. For example, the total amount of momentum and energy added by the winds is only a few percent (except in the highest mass-loss case). Meanwhile, the total energy and momentum inferred to be in the high velocity gas based on the CO emission is 20-30\%. As much as 50-60\% of the momentum and energy appears to be in the shells in the highest mass-loss run. This value is in fact comparable to the total fraction of cloud energy inferred to be in these shells for Perseus (50\%).

The discrepancy between the information obtained from the CO emission and 3D kinematic information,  Tables \ref{coprop} and \ref{bulkres}, respectively,  underscores that CO is a biased probe of the total underlying cloud momentum and energy. It preferentially selects optically thin gas with densities of $n_{\rm H_2}\simeq 500-5\times 10^3$ cm$^{-3}$,  and the average wind shell densities fall within this range.  The shells are also slightly warmer than the exterior gas, which enhances the CO excitation. We conclude that while CO emission is a useful diagnostic of cloud structure and may be used to identify the location and properties of shells, it likely over-estimates their impact compared to the 3D gas densities and velocities. 

The similarity between the simulated and observed CO emission leads to several practical conclusions. First, the mass-loss rates inferred from the observations are likely accurate; without energy injection of this magnitude it would be impossible to create the observed shell sizes and energies. Second, these mass-loss rates also reproduce the energy and momentum inferred from observed CO emission, which confirms that shell progenitors do play a significant role in shaping the gas morphology and energetics. Finally, since winds produce a clear signature in the molecular emission, evidence of this feature in observations is indicative of the presence of stellar winds.

 \subsection{Turbulent Driving}\label{driving}

A fundamental question motivating this study is ``Do stellar winds drive turbulence?" It is clear that significant gas motions do result as a product of the winds, but it is less clear whether these motions are purely local and, consequently, do not contribute much to the global turbulence. For compressible turbulent fluids, energy cascades from larger to smaller scales \citep[e.g.][]{sridhar94}, where the largest energy injection scale is the peak of the cascade. Little or no energy propagates from smaller to larger scales (a ``reverse" cascade).
 In the case of protostellar outflows, shocks created by the bipolar flows have a strong local impact, changing the fraction of solenoidal motions of the turbulence and creating new motions on small scales below the size of the outflow energy injection scale \citep[e.g., the linear extent of the outflow ][]{Offner14b}. Simulations show that within parsec sized clusters outflows can contribute sufficient turbulence to halt global collapse and regulate the star formation rate \citep{li07,carroll09,wang10,hansen12}. The dynamical timescale of the outflows is shorter than the global free fall time and it appears that outflows can replenish turbulence on scales that impact incipient star formation.  Likewise ionization from high-mass stars has a significant impact on cloud structure and appears to be able to inject sufficient energy to drive turbulence \citep{boneberg15}.  In contrast, the winds are largely spherical and sweep up gas as they advance. However, two effects limit the efficacy of energy injection by winds. First, while the energy injection scale, i.e., the effective radius of the bubble, increases with time, the bubble expansion stalls for mass-loss rates of $\lesssim 10^{-6} \msun$ yr$^{-1}$. This effect is enhanced by stronger magnetic fields. Second, high-velocity motions that are created are in the low-density material behind the shock, which remains inside the shells. This residual gas  contains very little mass and hence does not contribute much to the global turbulent cascade. 

\begin{figure}
\epsscale{1.3}
\vspace{-0.2in}
\plotone{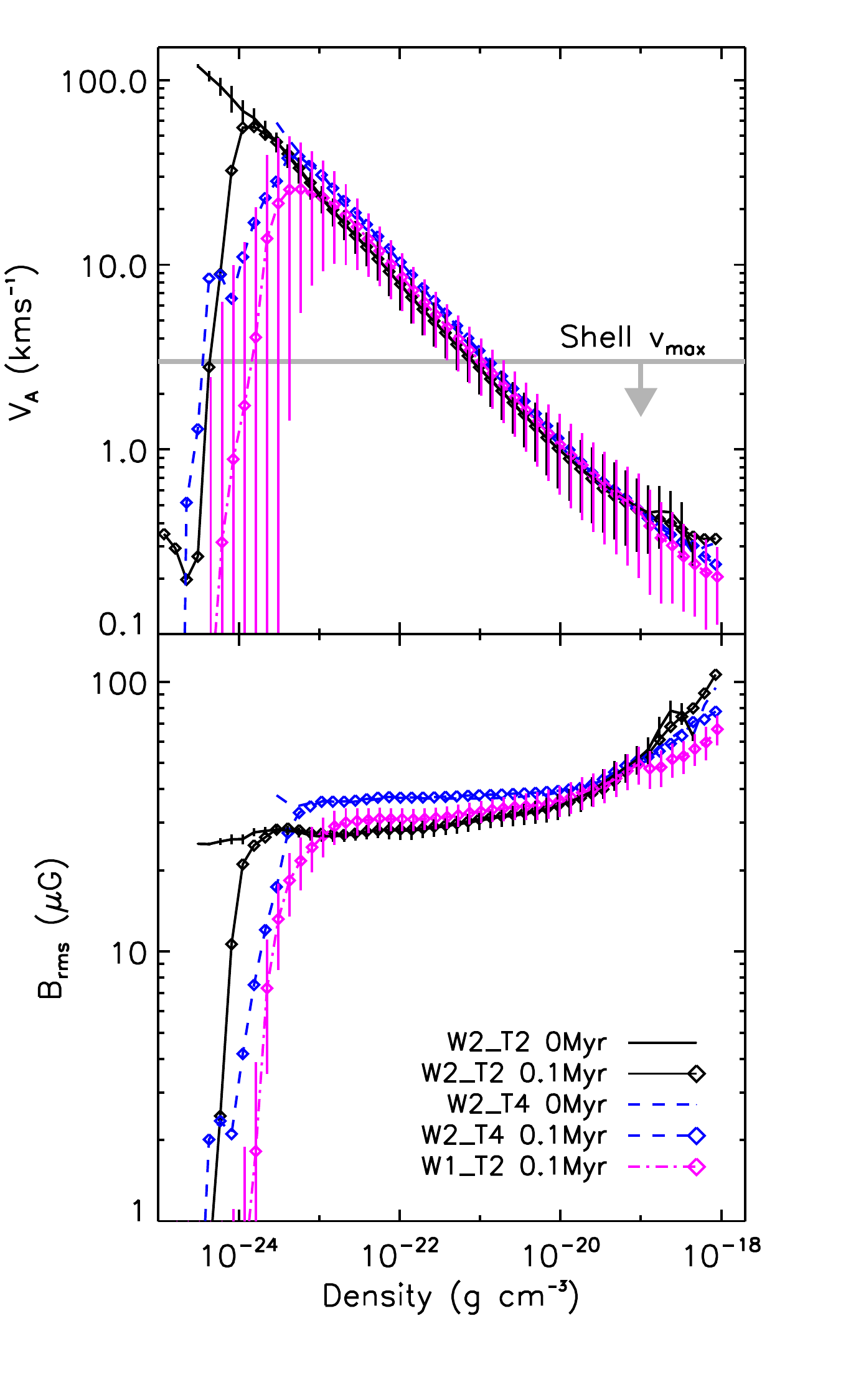}
\vspace{-0.4in}
\caption{Average Alf\'ven velocity (top) and average rms magnetic field strength (bottom) as a function of density. The averages are computed as a density weighted average over each bin. The error bars indicate $\pm 1 \sigma$ standard deviation in the binned values. For clarity, the dispersion is only shown for two representative cases; the standard deviation of the other wind and non-wind cases is similar.  The maximum shell expansion velocity is indicated by the grey solid line for 0.1 Myr. }\label{alfven}
\end{figure}

Whether winds can drive turbulence non-locally depends upon whether there is any impact on gas motions outside the shell radius. Alf\'ven waves may provide one avenue to excite non-local motions \citep[e.g.,][]{heitsch02}. The expanding shells displace the magnetic field, which excites Alf\'ven (transverse) waves that propagate along the field lines with a velocity $v_A= B/\sqrt{4 \pi \rho}$, where $B$ is the magnetic field strength and $\rho$ is the gas density. For the average simulation field strength and gas density, Alf\'ven waves  propagate with a velocity of 0.8 $\kms$, which is less than the shell expansion velocity of 1-5 $\kms$.   Fast magnetosonic waves  may also transport energy away from the shells   \citep{gendelev12}. Fast waves propagate perpendicular to the field with a speed $v_f = \sqrt{v_A^2+c_s^2}$, which is faster than the Alf\'ven speed but not significantly so since the sound speed in the gas just outside of the shells is $c_s\simeq 0.2~ \kms$. Due to the turbulent nature of the gas and tangled field morphology, it is not possible to directly see the propagation of magnetosonic waves as in \citet{gendelev12}. Figure \ref{alfven} shows the distribution of Al\'ven velocities with and without wind driving. The shocked shell gas ($\rho > 10^{-20}$ g cm$^{-3}$) has Alf\'ven velocities smaller than the shell expansion speed. Any excited magnetic waves may also rapidly damp \citep{banerjee07}. Consequently, we find the runs with winds only exhibit a minor enhancement in field strength and Alf\'ven speed at high densities. Little change occurs at intermediate and lower densities. The evacuated regions have densities $\lesssim 10^{-23}$ g cm$^{-3}$ and a relatively small magnetic field.  In summary, this suggests that shells do not contribute to the turbulent cascade except very locally.

\subsection{Scenarios for the Shell Origins}\label{altscenario}

Our simulations confirm the inferred mass-loss rates and wind momenta required to produce observed shells within low-mass star-forming molecular clouds. However, these results reinforce the severity of the weak wind problem and pose a significant challenge to our understanding of stellar feedback.  Here, we compare and evaluate explanations for these features.


\subsubsection{Line-Driven Winds}

 The primary explanation for these features, line-driven winds from main-sequence stars, is marked by a primary physical limitation. Photon driven winds are limited by the amount of momentum in the radiation field, which is set by the stellar luminosity, $L$. If all the photons radiated by the star are absorbed in the wind then the momentum injection rate is $\dot M_w v_w = L/c.$ This assumes that only the first scattered photon contributes to the radiation pressure (``the single scattering upper limit"). If multiple photon scatterings contribute then the momentum injection may be boosted by a factor of the optical depth of the wind, $\tau_w$, where $\tau_w$ is simply assumed to be constant with frequency. Thus, if the observed shells are created predominately by line-driven stellar mass-loss, then the mass-loss rate is \citep{lamers}:
\beqa
\dot M_w &=& \frac{\tau_w L}{v_w c} \nonumber \\
&=& 2\times10^{-6} \msun {\rm yr}^{-1} \left( \frac{\tau_w}{2} \right)\left( \frac{L}{10^4 \lsun} \right)  \nonumber \\
& & \times \left( \frac{v_w}{200 \kms} \right)^{-1},
\eeqa
where $10^4 ~\lsun$ is approximately the luminosity of a B0.5 star and the lines are assumed to be optically thick. This estimation is sufficient to account for the inferred mass-loss rate for the strongest source ($\dot m_w=2 \times 10^{-6} \msun$yr$^{-1}$) in both the W2 simulation and in Perseus. 

In practice, main-sequence winds are driven in narrow frequency bands, over which various atomic lines are optically thick. Higher luminosity stars have more saturated lines and hence can drive stronger winds, while dimmer stars have fewer or no optically thick lines for which momentum can be deposited. This is the essence of the weak-wind problem. Thus, the above estimate, which requires high-optical depth and multiple-scattering, is optimistic. These complications suggest that another factor or additional mechanism likely enhances momentum deposition in this classic line-driven wind picture.


\subsubsection{Wind Variability}

While the {\it average} wind velocities and stellar mass-loss may be orders of magnitude lower than we predict here, it is possible that stars, especially very young ones, experience periodic wind enhancements due to elevated magnetic stellar activity or accretion.  In fact, winds cannot be continuously three orders of magnitude higher than expected throughout the star lifetime since this would result in a significant change to the total stellar mass, e.g. of order $\sim \msun$.
However, given a mass-loss rate of $\gtrsim 10^{-7}\msun$ yr$^{-1}$, the simulations demonstrate that the shells are created relatively rapidly. This suggests that the observed features could plausibly be caused by intermittent high-activity or accretion over $\simeq 0.1$ Myr.  The cumulative affect of periodic wind enhancement could account for the shell sizes where a prolonged epoch of low mass-loss alone would be insufficient. 

Magnetic activity or accretion enhancements may be caused by interactions with a close stellar companion or interaction with some nearby residual gas or both \citep[e.g.,][]{basri97,jensen07}.  Since B-type stars do not have a convection zone in their outer atmosphere they are generally not expected to be magnetically active. However, instances of magnetically active B-stars have been observed \citep{gagne97, pillitteri14}. A sub-type of B stars, $\beta$-Cep variables, do experience X-ray pulsations, which may also lead to a wind enhancement \citep{oskinova14,oskinova15}. B-star $\rho$ Ophiuchi, which has produced a shell visible in the surrounding dust emission, is a binary {\it and} exhibits magnetic activity  \citep{pillitteri14}.

The multiplicity of the sources in Perseus  is currently unknown. Statistically, AB stars have a multiplicity fraction of 50-60\% \citep{duchene13}, which suggests that at least half of the wind progenitors may be in a binary system. However, the separation distribution peaks around 50 AU, so the fraction of wind progenitors with close or eccentric companions ($\lesssim10$AU) is considerably smaller.  

The gas distribution within the shells is also difficult to probe directly, since the gas is very low density and likely predominantly atomic.  HI and H$_2$ data does not exist at the resolution required to probe the intra-shell gas. Contamination from the foreground and background ISM also make distinguishing gas associated with the shell features difficult. However, further observations of the source environments and properties could be revealing. All these possibilities warrant follow-up studies of the wind progenitors in Perseus.

\subsubsection{Protostellar Outflows}

A significant amount of energy and momentum is deposited by magnetically collimated bipolar outflows during the first $\sim$0.5 Myr of the protostar stage. \citet{matzner00} estimate that the typical outflow momentum per unit stellar mass is $v_c = 40\kms$. This is an appropriate value for a region like Perseus which, given the number of observed protostars and a standard IMF, is likely not forming stars with masses above 3\msun \citep{matzner07,mckee10}. 
The momentum deposition rate of the outflows can then be written
\beq
\frac{dP_o}{dt} = \frac{m_*v_c}{\avg{t_p}},  
\eeq
where $m_*$ is the stellar mass and $\avg{t_p}$ is the typical protostellar lifetime, over which bipolar outflows occur.
The average rate of mass-loss due to outflows, $\dot m_o$,  can be written
\beqa
\dot m_o &=& \frac{m_*v_c}{t_p v_o} \\
&=&  2.4 \times 10^{-6}  \left(\frac{m_*}{3\msun} \right) \left( \frac{v_c}{40 \kms} \right) \left(\frac{t_p}{0.5 {\rm Myr}} \right)^{-1} \nonumber  \\
& & \times \left(\frac{v_o}{100 \kms} \right)^{-1},  \label{mdot}
\eeqa
where $v_o$ is the outflow launching velocity. Alternatively, given some instantaneous accretion rate for a particular source, the X-wind and disk-wind outflow models predict that the outflow rate will be a fixed fraction of the accretion rate,  $f_v=0.1$ and $f_v=0.33$, respectively \citep{shu88,pelletier92}.  Numerical models for disk wind launching find that the ratio of mass-loss to accretion rate can approach unity; however, there is an inverse correlation between mass loading and jet velocity, so this may not correspond to an increase in the total wind momentum \citep{sheik12}.  A  protostar accreting at $\dot m=10^{-5}\msun$yr$^{-1}$ would thus be expected to have an outflow rate of $\dot m_o \simeq 1-3 \times 10^{-6}\msun$yr$^{-1}$. The average accretion rate declines with time, so high-accretion is generally associated with the earliest stages of protostar formation, when the protostar is still embedded within a dense core \citep{dunham14}.

While these mass-loss rates are comparable to those inferred for observed shells, the highly collimated nature of protostellar outflows requires that the output energy and momentum decline steeply away from the outflow axis. \citet{matzner00} proposed that the force of the outflow deposition varies with angle according to
\beq
P(\theta) \simeq \frac{1}{ln(2/\theta_0)[1+\theta_0^2-{\rm cos}^2 (\theta)]},
\eeq
where $\theta$ is measured from the outflow axis and $\theta_0$ is the ``flattening angle" of the force distribution, over which the bulk of the momentum is injected. Through comparisons with observations, \citet{matzner00} conclude that $\theta_0 \lesssim 0.05$ and suggest a fiducial value of $\theta_0=0.01$. These values indicate that the momentum efficiency declines by a factor of 13-300 only $10^{\circ}$ from the outflow axis. 

If we suppose that the shell progenitors have a bipolar outflow pointed largely along the line-of-sight, we can evaluate whether the transverse component of the outflow would be sufficient to create the observed shells. For an outflow with an axis oriented $20^{\circ}$ from the line-of-sight, any motions of the gas in the plane of the sky would be caused by outflow material $70^{\circ}$ from the outflow axis. At this angle assuming a conservative amount of collimation, $\theta_0=0.05$, the outflow force would be diminished by a factor of $f_{\theta}=P(0^{\circ})/P(70^{\circ}) \simeq 350$. For a $3\msun$ protostar, this corresponds to an effective momentum deposition rate of $\dot P_o= f_{\theta} \dot m_o v_o \simeq 1/350 \times (2.4 \times 10^{-6} \msun$yr$^{-1}) \times $100$\kms \sim 5 \times 10^{-7} \msun$yr$^{-1}\kms$  (see Equation \ref{mdot}). This amount of momentum deposition is significantly less than the momentum injection by a 3$\msun$ star according to the \citet{vink01} models: $\dot P_w = \dot m_w v_w = 10^{-8}\msun$yr$^{-1} 600 \kms = 6 \times 10^{-6}  \msun$yr$^{-1}\kms$. Runs W3\_T2 and W4\_T2, which modeled isolated stars with these properties, demonstrated that this momentum is {\it too low} to create structures of the observed size and energy.

In order for off-axis magnetically collimated outflows to create shell features, protostellar  {\it outflow} rates would have to be tens to hundreds of times higher than those observed, i.e., $\dot m_o \gtrsim 10^{-4} \msun$yr$^{-1}$. Outflows of this magnitude are wildly discrepant with inferred protostellar accretion rates \citep{Offner11b,dunham14}, and there are no protostellar outflows identified within Perseus  that have this amount of mass flux \citep{arce10,arce11}. We conclude that protostellar outflows are very unlikely to produce the observed shells.



%

\subsubsection{Stellar Groups}

Small groups of stars may be able to provide sufficient integrated momentum to create a shell where a single star is insufficient.  For example, \citet{nakamura12} conjecture that the shells near the L1641-N and V 380 Ori clusters were formed by the combined winds (or protostellar outflow activity) of the cluster. Similarly, the shell surrounding rho Ophiuchi contains several additional stars that may contribute to the total momentum (I. Pillitteri priv. comm.). 
However, this still presents an incomplete solution. The most massive star in L1641-N is a B4 V Herbig Ae/Be star, which has an extrapolated mass-loss rate  $<10^{-9} \msun$ yr$^{-1}$.  The cluster contains 80 young stellar objects, but these cumulatively are unlikely to account for the needed momentum from stellar winds, even extrapolating the \citet{vink01} prescription to lower masses.  

Although stellar groups may account for the properties of other observed shells, those in Perseus do not appear to contain star clusters. Shell progenitors are at most binary or triple systems. It is possible future surveys of stellar kinematics may reveal other associated stars, which were previously overlooked.  

\subsubsection{Voids}

If voids are created by chance fluctuations in the turbulence then it is possible that low-density, coherent regions occasionally appear within clouds which are not due to feedback. If the gas is sufficiently low density then CO may not be present, i.e., the cloud is porous and the molecules are dissociated,  or the emission may not be detectible. Thus, voids could masquerade as wind features. Our synthetic CO maps do show a couple of regions without CO emission that are not created by stellar winds. However, enhanced CO on the boundary of these voids is hard to produce without a central source. The simulations don't provide evidence that chance alignments are likely to create similarly bright morphological features. Consequently, we conclude that voids created by turbulence are not likely to explain the majority of observed features, most of which do show a progenitor with high confidence. A few of the observed shells also display evidence of interior ionized gas and warm dust, which requires a bright source.

\section{Conclusions}\label{conclusions}

We use magnetohydrodynamic simulations to investigate the impact of stellar winds on molecular cloud structure and kinematics. We compute the $^{12}$CO(1-0) cloud emission and compare it with observed data. We demonstrate that the morphology of the resulting simulated shells is similar to structures observed in nearby clouds, which are attributed to spherical mass-loss from young stars. We explore a range of wind mass-loss rates and conclude that the observed size of the shells and their properties can only be created by progenitors with mass-loss rates of $10^{-7}-10^{-6} \msun$yr$^{-1}$, which is consistent with the rates estimated from the observed CO molecular line data. This highlights a discrepancy of several orders of magnitude between observations and models of wind launching, which predict mass-loss rates of $<10^{-8}\msun$yr$^{-1}$ for the stellar types identified.  

 We evaluate the ability of the winds to drive supersonic turbulence on scales of a few parsecs. 
 A significant contribution to the turbulent cascade from winds hinges on the creation of non-local perturbations; otherwise the winds only deposit energy and momentum in the shells and do not offset global turbulent dissipation.  While we find that the winds do create significant gas velocity enhancements compared to the initial cloud turbulence, which are especially notable in CO line emission,  this appears to be a purely localized effect. Expansion slows considerably after the first 0.05Myr and most shells become pressure confined within a few 0.1Myr. We find that the winds leave no signatures in the density and momentum distribution functions. The winds do create a feature at $\sim 1$pc in the velocity Fourier spectrum, but no corresponding feature appears in the power spectra of the integrated CO emission.  We find no evidence, even in the strongest wind case, that the winds offset the global turbulent dissipation. 
 
 Despite the localized nature of the wind deposition, we demonstrate that the energy and momentum of the shells can be considerable compared to the total cloud energy and momentum. However, energetics estimated using CO emission, which selects a particular range of intermediate gas densities, tends to over-estimate the true wind impact. The high-velocity component ($v>1\kms$) of the CO emission suggests that 10-50\% of the total cloud energy and momentum resides in shells. However, with the exception of the strongest wind case, the true contribution as measured using all the 3D gas information is not more than $\sim$5\%.
 
 We discuss alternative proposed origins for the shells, including turbulent voids, protostellar outflows, and wind variability, and we conclude that only variability presents a plausible alternative explanation. These results have significant implications for our understanding of stellar winds, and our work suggests wind mass-loss rates are significantly underestimated by current models.  While the impact of winds from AB stars remains smaller than those of more massive O stars, AB stars are significantly more numerous and could provide significant impact in the aggregate. Consequently, underestimates of mass-loss rates have implications for molecular cloud properties and galaxy evolution within cosmological simulations \citep[e.g.,][]{agertz14}. Future numerical work including gravity will examine the direct impact of wind feedback on star formation activity. 
  
\acknowledgements 
 Thanks to Eve Ostriker, Christoph Federrath, Suzan Edwards, Jim Dale, Martin Weinberg, Nathan Smith, and Ignazio Pillitteri for helpful discussions. We also thank the anonymous referee for comments that improved this manuscript.   The data analysis, images and animations were made possible by {\it yt} \citep{turk11}.
Partial support for S.S.R.O. was provided by NASA through Hubble Fellowship grant HF-51311.01 
awarded by the Space Telescope Science Institute, which is operated by the Association of
Universities for Research in Astronomy, Inc., for NASA, under contract NAS 5-26555. 
H.G.A. acknowledges support from his NSF CAREER award AST-0845619. The simulations were performed on the Yale University Omega cluster; this work was supported in part by the facilities and staff of the Yale University Faculty of Arts and Sciences High Performance Computing Center.

\appendix

\section{Analytic Comparison of Wind Expansion}\label{appendix1}

We test the wind implementation by computing the expansion of the wind launched by a single source embedded in a uniform density medium.  We note the {\sc orion} hydrodynamic methodology has been benchmarked against a variety of standard tests in prior work \citep[e.g.,][]{li12}.  

\begin{figure}
\epsscale{0.75}
\vspace{-0.1in}
\plotone{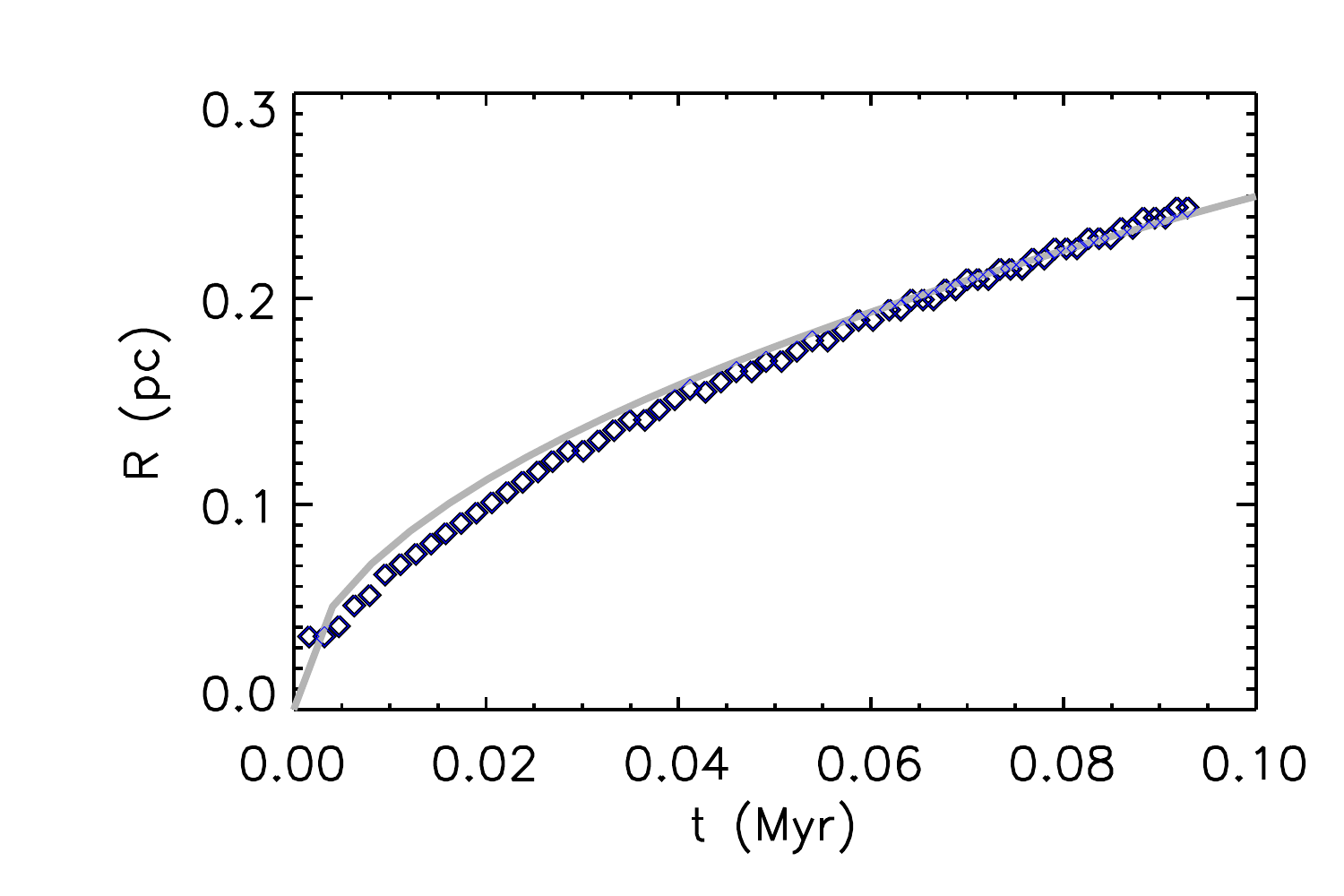}
\vspace{-0.1in}
\caption{ Radius of the shell shock front as a function of time (blue diamonds) with ambient density $\rho_0 =2.04 \times 10^{-21}$gcm$^{-3}$ (the fiducial cloud density in the models) and a source with mass-loss rate $\dot m_w = 6\times 10^{-8}\msun$yr$^{-1}$. The grey solid line gives the corresponding analytic solution from equation \ref{KMRs}. } \label{expansion}
\end{figure}

 \citet{koo92}, henceforth KM92, have solved for the expansion of wind bubbles in a uniform medium given some constant mass-loss rate. KM92 write the mass-loss rate, $\dot M_{\rm w}$, in terms of the mechanical luminosity, $L$,  as $\dot  M_{\rm in}  = 2L/v_{\rm w}^2$, where $v_w$ is the wind velocity. We rewrite and renormalize their solutions here in terms of $\dot M_{w}$. The size of an expanding radiative shell as a function of time is then (KM92, equation 3.1):
\beq
R_s(t) = 0.24 \left( \frac{\dot M_w}{10^{-7} \msun {\rm yr}^{-1}} \right)^{1/4} \left( \frac{v_w}{200 \kms} \right)^{1/4} \left( \frac{n_{\rm H}}{872 {\rm cm}^{-3} } \right)^{-1/4} \left( \frac{t}{ \rm 0.1Myr} \right)^{1/2} \rm{pc}, \label{KMRs}
\eeq
where $n_{\rm H} = \rho_0/2.34\times 10^{-24}$g is the Hydrogen number density.
Figure \ref{expansion} shows good agreement between the location of the wind shock front and the expected analytic solution. 

\section{Wind Expansion in a Turbulent Magnetized Medium}\label{appendix2}

Although the gas in molecular clouds is non-uniform due to turbulence, the KM92 solutions also apply to the average expansion with some modification. Since the winds here expand into a cold ISM, the swept-up shell material is also relatively cold (the entrained gas mass is much larger than the mass in the wind) and the winds can be described by the slow-wind (radiative) solution.

Once the ram pressure of the shell is equal to the ambient pressure, the shell becomes ``pressure-confined" and the expansion stops. The time that this occurs for a radiative bubble is (KM92, equation 3.8):
\beq
t_P =  3.5 \left( \frac{\dot M_w}{10^{-7} \msun {\rm yr}^{-1}} \right)^{1/2} \left( \frac{v_w}{200 \kms} \right)^{-3/2} \left( \frac{n_{\rm H}}{872 {\rm cm}^{-3}}  \right) ^{-1/2} \left( \frac{\mathcal{M}}{ 1000} \right)^{2} \rm{Myr},
\eeq
where $\mathcal{M}_w = v_w/c_s$ is the effective Mach number of the wind as compared to the isothermal sound speed in the ambient medium. Given a wind velocity of 200 $\kms$ and a gas temperature of 10 K, $c_s \simeq 0.2\kms$ and $\mathcal{M}_w = 1000$. This predicts that a wind with $10^{-7} \msun$ yr$^{-1}$ would not become pressure confined for more than 3 Myr. However, the ambient medium here is both magnetized and turbulent. Consequently, it is appropriate to replace $c_s$ with the effective sound speed, $c_{\rm s,eff} = \sqrt{c_s^2+\sigma_{\rm 1D}^2 + v_{A}^2}$, where $\sigma_{\rm 1D}$ is the rms 1D turbulent velocity dispersion and $v_A$ is the Alf\'ven speed. For our fiducial simulation, $c_{\rm eff} = 1.43 \kms$, which means that a source with a mass-loss rate $10^{-7} \msun {\rm yr}^{-1}$ will become pressure confined in $t_P=0.068$ Myr.

\section{Fourier Spectra Resolution Comparison}\label{appendix3}

\begin{figure}
\epsscale{1.1}
\vspace{-0.1in}
\plottwo{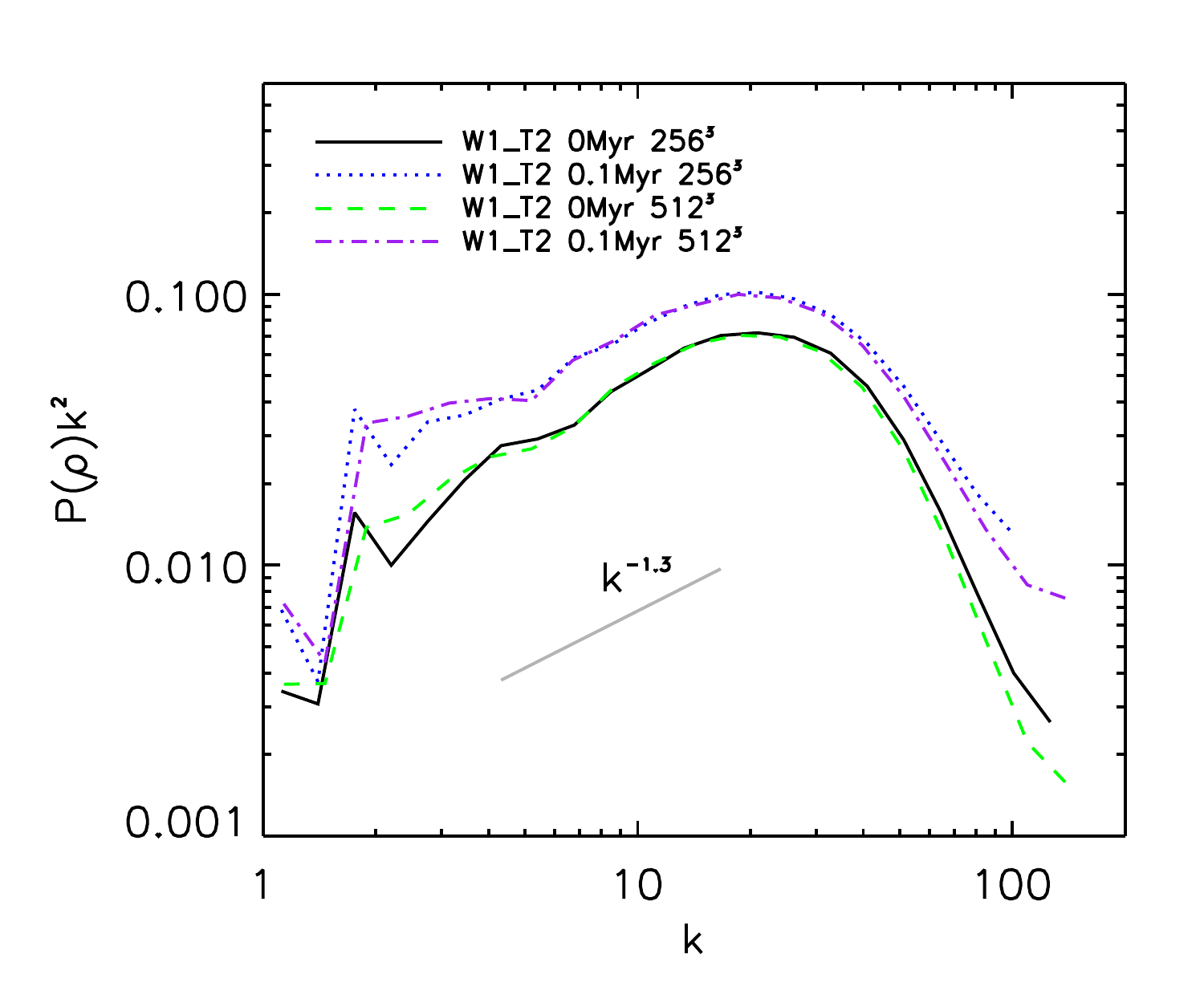}{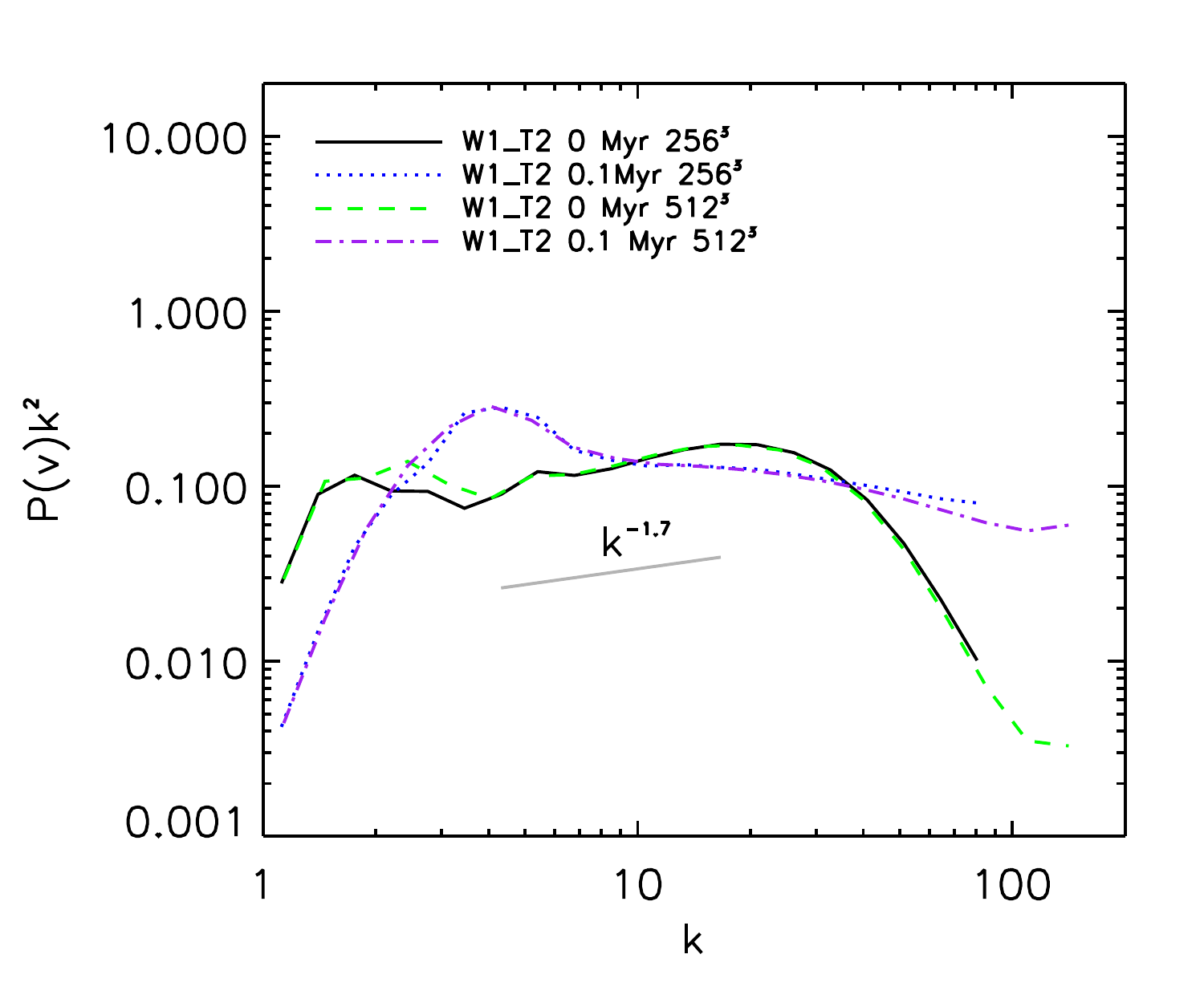}
\vspace{-0.1in}
\caption{Log density (left) and velocity (right) Fourier spectra computed using 256$^3$ and 512$^3$ grid resolutions for two different times. Power laws of $k^{-1.3}$  and $k^{-1.7}$, respectively, and approximate inertial range are denoted by grey lines. The spectra have been compensated by $k^2$.}\label{power_res}
\end{figure}

The basegrid for the calculation that is used to generate the initial turbulent conditions is $256^3$.  Larger grid sizes are generally preferred for computing turbulent statistics \citep{kritsuk07}. Figure \ref{power_res} compares the density power spectrum for the 256$^3$ data and for a flattened grid of $512^3$ that takes into account the AMR level 1 data. In the latter case, strong shocks as well as the wind shells are captured at finer resolution. The power law exponents of the lower resolution spectra are $\sim 0.1$ flatter than the higher resolution power laws, where the higher resolution is more similar to prior published values for turbulent calculations. The 256$^3$ resolution momentum and velocity Fourier spectra power law fits are also flatter by $\sim 0.1$.

\bibliography{biblio}

\begin{thebibliography}{}
\expandafter\ifx\csname natexlab\endcsname\relax\def\natexlab#1{#1}\fi

\bibitem[{{Agertz} \& {Kravtsov}(2014)}]{agertz14}
{Agertz}, O., \& {Kravtsov}, A.~V. 2014, ArXiv e-prints, arXiv:1404.2613

\bibitem[{{Arce} {et~al.}(2011){Arce}, {Borkin}, {Goodman}, {Pineda}, \&
  {Beaumont}}]{arce11}
{Arce}, H.~G., {Borkin}, M.~A., {Goodman}, A.~A., {Pineda}, J.~E., \&
  {Beaumont}, C.~N. 2011, \apj, 742, 105

\bibitem[{{Arce} {et~al.}(2010){Arce}, {Borkin}, {Goodman}, {Pineda}, \&
  {Halle}}]{arce10}
{Arce}, H.~G., {Borkin}, M.~A., {Goodman}, A.~A., {Pineda}, J.~E., \& {Halle},
  M.~W. 2010, \apj, 715, 1170

\bibitem[{{Arthur} {et~al.}(2011){Arthur}, {Henney}, {Mellema}, {de Colle}, \&
  {V{\'a}zquez-Semadeni}}]{arthur11}
{Arthur}, S.~J., {Henney}, W.~J., {Mellema}, G., {de Colle}, F., \&
  {V{\'a}zquez-Semadeni}, E. 2011, \mnras, 414, 1747

\bibitem[{{Banerjee} \& {Pudritz}(2007)}]{banerjee07}
{Banerjee}, R., \& {Pudritz}, R.~E. 2007, \apj, 660, 479

\bibitem[{{Basri} {et~al.}(1997){Basri}, {Johns-Krull}, \& {Mathieu}}]{basri97}
{Basri}, G., {Johns-Krull}, C.~M., \& {Mathieu}, R.~D. 1997, \aj, 114, 781

\bibitem[{{Beaumont} {et~al.}(2013){Beaumont}, {Offner}, {Shetty}, {Glover}, \&
  {Goodman}}]{beaumont13}
{Beaumont}, C.~N., {Offner}, S.~S.~R., {Shetty}, R., {Glover}, S.~C.~O., \&
  {Goodman}, A.~A. 2013, \apj, 777, 173

\bibitem[{{Beaumont} \& {Williams}(2010)}]{beaumont10}
{Beaumont}, C.~N., \& {Williams}, J.~P. 2010, \apj, 709, 791

\bibitem[{{Boneberg} {et~al.}(2015){Boneberg}, {Dale}, {Girichidis}, \&
  {Ercolano}}]{boneberg15}
{Boneberg}, D.~M., {Dale}, J.~E., {Girichidis}, P., \& {Ercolano}, B. 2015,
  \mnras, 447, 1341

\bibitem[{{Burgers}(1948)}]{burgers48}
{Burgers}, J.~M. 1948, Adv. Appl. Mech., 1, 171

\bibitem[{{Burkhart} {et~al.}(2013){Burkhart}, {Lazarian}, {Ossenkopf}, \&
  {Stutzki}}]{burkhart13}
{Burkhart}, B., {Lazarian}, A., {Ossenkopf}, V., \& {Stutzki}, J. 2013, \apj,
  771, 123

\bibitem[{{Burkhart} {et~al.}(2010){Burkhart}, {Stanimirovi{\'c}}, {Lazarian},
  \& {Kowal}}]{burkhart10}
{Burkhart}, B., {Stanimirovi{\'c}}, S., {Lazarian}, A., \& {Kowal}, G. 2010,
  \apj, 708, 1204

\bibitem[{{Carroll} {et~al.}(2010){Carroll}, {Frank}, \&
  {Blackman}}]{carroll10}
{Carroll}, J.~J., {Frank}, A., \& {Blackman}, E.~G. 2010, \apj, 722, 145

\bibitem[{{Carroll} {et~al.}(2009){Carroll}, {Frank}, {Blackman}, {Cunningham},
  \& {Quillen}}]{carroll09}
{Carroll}, J.~J., {Frank}, A., {Blackman}, E.~G., {Cunningham}, A.~J., \&
  {Quillen}, A.~C. 2009, \apj, 695, 1376

\bibitem[{{Cunningham} {et~al.}(2011){Cunningham}, {Klein}, {Krumholz}, \&
  {McKee}}]{cunningham11}
{Cunningham}, A.~J., {Klein}, R.~I., {Krumholz}, M.~R., \& {McKee}, C.~F. 2011,
  \apj, 740, 107

\bibitem[{{Dale} \& {Bonnell}(2008)}]{dale08}
{Dale}, J.~E., \& {Bonnell}, I.~A. 2008, \mnras, 391, 2

\bibitem[{{Dale} {et~al.}(2013){Dale}, {Ngoumou}, {Ercolano}, \&
  {Bonnell}}]{dale13}
{Dale}, J.~E., {Ngoumou}, J., {Ercolano}, B., \& {Bonnell}, I.~A. 2013, \mnras,
  436, 3430

\bibitem[{{Deharveng} {et~al.}(2010){Deharveng}, {Schuller}, {Anderson},
  {Zavagno}, {Wyrowski}, {Menten}, {Bronfman}, {Testi}, {Walmsley}, \&
  {Wienen}}]{deharveng10}
{Deharveng}, L., {Schuller}, F., {Anderson}, L.~D., {et~al.} 2010, \aap, 523,
  A6

\bibitem[{{Duch{\^e}ne} \& {Kraus}(2013)}]{duchene13}
{Duch{\^e}ne}, G., \& {Kraus}, A. 2013, \araa, 51, 269

\bibitem[{{Dunham} {et~al.}(2014){Dunham}, {Stutz}, {Allen}, {Evans},
  {Fischer}, {Megeath}, {Myers}, {Offner}, {Poteet}, {Tobin}, \&
  {Vorobyov}}]{dunham14}
{Dunham}, M.~M., {Stutz}, A.~M., {Allen}, L.~E., {et~al.} 2014, Protostars and
  Planets VI, 195

\bibitem[{{Federrath}(2013)}]{federrath13}
{Federrath}, C. 2013, \mnras, 436, 1245

\bibitem[{{Federrath} {et~al.}(2008){Federrath}, {Klessen}, \&
  {Schmidt}}]{federrath08}
{Federrath}, C., {Klessen}, R.~S., \& {Schmidt}, W. 2008, \apjl, 688, L79

\bibitem[{{Federrath} {et~al.}(2010){Federrath}, {Roman-Duval}, {Klessen},
  {Schmidt}, \& {Mac Low}}]{federrath10}
{Federrath}, C., {Roman-Duval}, J., {Klessen}, R.~S., {Schmidt}, W., \& {Mac
  Low}, M.-M. 2010, \aap, 512, A81

\bibitem[{{Foster} {et~al.}(2015){Foster}, {Cottaar}, {Covey}, {Arce}, {Meyer},
  {Nidever}, {Stassun}, {Tan}, {Chojnowski}, {da Rio}, {Flaherty}, {Rebull},
  {Frinchaboy}, {Majewski}, {Skrutskie}, {Wilson}, \& {Zasowski}}]{foster15}
{Foster}, J.~B., {Cottaar}, M., {Covey}, K.~R., {et~al.} 2015, \apj, 799, 136

\bibitem[{{Frank} {et~al.}(2014){Frank}, {Ray}, {Cabrit}, {Hartigan}, {Arce},
  {Bacciotti}, {Bally}, {Benisty}, {Eisl{\"o}ffel}, {G{\"u}del}, {Lebedev},
  {Nisini}, \& {Raga}}]{frankppvi14}
{Frank}, A., {Ray}, T.~P., {Cabrit}, S., {et~al.} 2014, Protostars and Planets
  VI, 451

\bibitem[{{Frerking} {et~al.}(1982){Frerking}, {Langer}, \&
  {Wilson}}]{frerking82}
{Frerking}, M.~A., {Langer}, W.~D., \& {Wilson}, R.~W. 1982, \apj, 262, 590

\bibitem[{{Gagn{\'e}} {et~al.}(1997){Gagn{\'e}}, {Caillault}, {Stauffer}, \&
  {Linsky}}]{gagne97}
{Gagn{\'e}}, M., {Caillault}, J.-P., {Stauffer}, J.~R., \& {Linsky}, J.~L.
  1997, \apjl, 478, L87

\bibitem[{{Geen} {et~al.}(2015){Geen}, {Rosdahl}, {Blaizot}, {Devriendt}, \&
  {Slyz}}]{geen15}
{Geen}, S., {Rosdahl}, J., {Blaizot}, J., {Devriendt}, J., \& {Slyz}, A. 2015,
  \mnras, 448, 3248

\bibitem[{{Gendelev} \& {Krumholz}(2012)}]{gendelev12}
{Gendelev}, L., \& {Krumholz}, M.~R. 2012, \apj, 745, 158

\bibitem[{{Hansen} {et~al.}(2012){Hansen}, {Klein}, {McKee}, \&
  {Fisher}}]{hansen12}
{Hansen}, C.~E., {Klein}, R.~I., {McKee}, C.~F., \& {Fisher}, R.~T. 2012, \apj,
  747, 22

\bibitem[{{Heitsch} \& {Burkert}(2002)}]{heitsch02}
{Heitsch}, F., \& {Burkert}, A. 2002, in Astronomical Society of the Pacific
  Conference Series, Vol. 285, Modes of Star Formation and the Origin of Field
  Populations, ed. E.~K. {Grebel} \& W.~{Brandner}, 13

\bibitem[{{Jensen} {et~al.}(2007){Jensen}, {Dhital}, {Stassun}, {Patience},
  {Herbst}, {Walter}, {Simon}, \& {Basri}}]{jensen07}
{Jensen}, E.~L.~N., {Dhital}, S., {Stassun}, K.~G., {et~al.} 2007, \aj, 134,
  241

\bibitem[{{Kolmogorov}(1941)}]{kolmogorov41}
{Kolmogorov}, A.~N. 1941, Akademiia Nauk SSSR Doklady, 32, 16

\bibitem[{{Koo} \& {McKee}(1992)}]{koo92}
{Koo}, B.-C., \& {McKee}, C.~F. 1992, \apj, 388, 93

\bibitem[{{Kritsuk} {et~al.}(2007){Kritsuk}, {Norman}, {Padoan}, \&
  {Wagner}}]{kritsuk07}
{Kritsuk}, A.~G., {Norman}, M.~L., {Padoan}, P., \& {Wagner}, R. 2007, \apj,
  665, 416

\bibitem[{{Krumholz} {et~al.}(2007){Krumholz}, {Klein}, {McKee}, \&
  {Bolstad}}]{krumholzkmb07}
{Krumholz}, M.~R., {Klein}, R.~I., {McKee}, C.~F., \& {Bolstad}, J. 2007, \apj,
  667, 626

\bibitem[{{Krumholz} {et~al.}(2009){Krumholz}, {Klein}, {McKee}, {Offner}, \&
  {Cunningham}}]{krumholz09}
{Krumholz}, M.~R., {Klein}, R.~I., {McKee}, C.~F., {Offner}, S.~S.~R., \&
  {Cunningham}, A.~J. 2009, Science, 323, 754

\bibitem[{{Krumholz} {et~al.}(2004){Krumholz}, {McKee}, \&
  {Klein}}]{krumholz04}
{Krumholz}, M.~R., {McKee}, C.~F., \& {Klein}, R.~I. 2004, \apj, 611, 399

\bibitem[{{Lamers} \& {Cassinelli}(1999)}]{lamers}
{Lamers}, H.~J.~G.~L.~M., \& {Cassinelli}, J.~P. 1999, {Introduction to Stellar
  Winds}

\bibitem[{{Lamers} {et~al.}(1995){Lamers}, {Snow}, \& {Lindholm}}]{lamers95}
{Lamers}, H.~J.~G.~L.~M., {Snow}, T.~P., \& {Lindholm}, D.~M. 1995, \apj, 455,
  269

\bibitem[{{Lazarian} \& {Pogosyan}(2004)}]{lazarian04}
{Lazarian}, A., \& {Pogosyan}, D. 2004, \apj, 616, 943

\bibitem[{{Lee} {et~al.}(2014){Lee}, {Cunningham}, {McKee}, \& {Klein}}]{lee14}
{Lee}, A.~T., {Cunningham}, A.~J., {McKee}, C.~F., \& {Klein}, R.~I. 2014,
  \apj, 783, 50

\bibitem[{{Lee} {et~al.}(2015){Lee}, {Hirano}, {Zhang}, {Shang}, {Ho}, \&
  {Mizuno}}]{lee15}
{Lee}, C.-F., {Hirano}, N., {Zhang}, Q., {et~al.} 2015, \apj, 805, 186

\bibitem[{{Li} {et~al.}(2012){Li}, {Martin}, {Klein}, \& {McKee}}]{li12}
{Li}, P.~S., {Martin}, D.~F., {Klein}, R.~I., \& {McKee}, C.~F. 2012, \apj,
  745, 139

\bibitem[{{Mac Low}(1999)}]{maclow99}
{Mac Low}, M.-M. 1999, \apj, 524, 169

\bibitem[{{Matzner}(2007)}]{matzner07}
{Matzner}, C.~D. 2007, \apj, 659, 1394

\bibitem[{{Matzner} \& {McKee}(2000)}]{matzner00}
{Matzner}, C.~D., \& {McKee}, C.~F. 2000, \apj, 545, 364

\bibitem[{{McKee}(1989)}]{mckee89}
{McKee}, C.~F. 1989, \apj, 345, 782

\bibitem[{{McKee} \& {Offner}(2010)}]{mckee10}
{McKee}, C.~F., \& {Offner}, S.~S.~R. 2010, \apj, 716, 167

\bibitem[{{McKee} \& {Ostriker}(2007)}]{MandO07}
{McKee}, C.~F., \& {Ostriker}, E. 2007, \araa, 45, 565

\bibitem[{{Mignone} {et~al.}(2007){Mignone}, {Bodo}, {Massaglia}, {Matsakos},
  {Tesileanu}, {Zanni}, \& {Ferrari}}]{mignone07}
{Mignone}, A., {Bodo}, G., {Massaglia}, S., {et~al.} 2007, \apjs, 170, 228

\bibitem[{{Mignone} {et~al.}(2012){Mignone}, {Zanni}, {Tzeferacos}, {van
  Straalen}, {Colella}, \& {Bodo}}]{mignone12}
{Mignone}, A., {Zanni}, C., {Tzeferacos}, P., {et~al.} 2012, \apjs, 198, 7

\bibitem[{{Miniati} \& {Colella}(2007)}]{miniati07}
{Miniati}, F., \& {Colella}, P. 2007, Journal of Computational Physics, 227,
  400

\bibitem[{{Myers} {et~al.}(2014){Myers}, {Klein}, {Krumholz}, \&
  {McKee}}]{myers14}
{Myers}, A.~T., {Klein}, R.~I., {Krumholz}, M.~R., \& {McKee}, C.~F. 2014,
  \mnras, 439, 3420

\bibitem[{{Nakamura} \& {Li}(2007)}]{li07}
{Nakamura}, F., \& {Li}, Z.-Y. 2007, \apj, 662, 395

\bibitem[{{Nakamura} {et~al.}(2012){Nakamura}, {Miura}, {Kitamura},
  {Shimajiri}, {Kawabe}, {Akashi}, {Ikeda}, {Tsukagoshi}, {Momose}, {Nishi}, \&
  {Li}}]{nakamura12}
{Nakamura}, F., {Miura}, T., {Kitamura}, Y., {et~al.} 2012, \apj, 746, 25

\bibitem[{{Ntormousi} {et~al.}(2011){Ntormousi}, {Burkert}, {Fierlinger}, \&
  {Heitsch}}]{ntormousi11}
{Ntormousi}, E., {Burkert}, A., {Fierlinger}, K., \& {Heitsch}, F. 2011, \apj,
  731, 13

\bibitem[{{Offner} \& {Arce}(2014)}]{Offner14b}
{Offner}, S.~S.~R., \& {Arce}, H.~G. 2014, \apj, 784, 61

\bibitem[{{Offner} {et~al.}(2014){Offner}, {Clark}, {Hennebelle}, {Bastian},
  {Bate}, {Hopkins}, {Moraux}, \& {Whitworth}}]{Offnerppvi}
{Offner}, S.~S.~R., {Clark}, P.~C., {Hennebelle}, P., {et~al.} 2014, Protostars
  and Planets VI, 53

\bibitem[{{Offner} {et~al.}(2009){Offner}, {Klein}, {McKee}, \&
  {Krumholz}}]{Offner09c}
{Offner}, S.~S.~R., {Klein}, R.~I., {McKee}, C.~F., \& {Krumholz}, M.~R. 2009,
  \apj, 703, 131

\bibitem[{{Offner} \& {McKee}(2011)}]{Offner11b}
{Offner}, S.~S.~R., \& {McKee}, C.~F. 2011, \apj, 736, 53

\bibitem[{{Oskinova} {et~al.}(2014){Oskinova}, {Naz{\'e}}, {Todt},
  {Huenemoerder}, {Ignace}, {Hubrig}, \& {Hamann}}]{oskinova14}
{Oskinova}, L.~M., {Naz{\'e}}, Y., {Todt}, H., {et~al.} 2014, Nature
  Communications, 5, 4024

\bibitem[{{Oskinova} {et~al.}(2015){Oskinova}, {Todt}, {Huenemoerder},
  {Hubrig}, {Ignace}, {Hamann}, \& {Balona}}]{oskinova15}
{Oskinova}, L.~M., {Todt}, H., {Huenemoerder}, D.~P., {et~al.} 2015, \aap, 577,
  A32

\bibitem[{{Owocki}(2000)}]{owocki00}
{Owocki}, S. 2000, {Radiatively Driven Stellar Winds from Hot Stars}, ed.
  P.~{Murdin}, 1887

\bibitem[{{Padoan} \& {Nordlund}(2004)}]{padoan04}
{Padoan}, P., \& {Nordlund}, {\AA}. 2004, \apj, 617, 559

\bibitem[{{Pelletier} \& {Pudritz}(1992)}]{pelletier92}
{Pelletier}, G., \& {Pudritz}, R.~E. 1992, \apj, 394, 117

\bibitem[{{Pillitteri} {et~al.}(2014){Pillitteri}, {Wolk}, {Goodman}, \&
  {Sciortino}}]{pillitteri14}
{Pillitteri}, I., {Wolk}, S.~J., {Goodman}, A., \& {Sciortino}, S. 2014, \aap,
  567, L4

\bibitem[{{Ridge} {et~al.}(2006{\natexlab{a}}){Ridge}, {Schnee}, {Goodman}, \&
  {Foster}}]{ridge06b}
{Ridge}, N.~A., {Schnee}, S.~L., {Goodman}, A.~A., \& {Foster}, J.~B.
  2006{\natexlab{a}}, \apj, 643, 932

\bibitem[{{Ridge} {et~al.}(2006{\natexlab{b}}){Ridge}, {Di Francesco}, {Kirk},
  {Li}, {Goodman}, {Alves}, {Arce}, {Borkin}, {Caselli}, {Foster}, {Heyer},
  {Johnstone}, {Kosslyn}, {Lombardi}, {Pineda}, {Schnee}, \&
  {Tafalla}}]{ridge06}
{Ridge}, N.~A., {Di Francesco}, J., {Kirk}, H., {et~al.} 2006{\natexlab{b}},
  \aj, 131, 2921

\bibitem[{{Rogers} \& {Pittard}(2013)}]{rogers13}
{Rogers}, H., \& {Pittard}, J.~M. 2013, \mnras, 431, 1337

\bibitem[{{Rogers} \& {Pittard}(2014)}]{rogers14}
---. 2014, \mnras, 441, 964

\bibitem[{{Sheikhnezami} {et~al.}(2012){Sheikhnezami}, {Fendt}, {Porth},
  {Vaidya}, \& {Ghanbari}}]{sheik12}
{Sheikhnezami}, S., {Fendt}, C., {Porth}, O., {Vaidya}, B., \& {Ghanbari}, J.
  2012, \apj, 757, 65

\bibitem[{{Shetty} {et~al.}(2011){Shetty}, {Glover}, {Dullemond}, \&
  {Klessen}}]{shetty11}
{Shetty}, R., {Glover}, S.~C., {Dullemond}, C.~P., \& {Klessen}, R.~S. 2011,
  \mnras, 412, 1686

\bibitem[{{Shu}(1992)}]{shutextbook}
{Shu}, F.~H. 1992, {Physics of Astrophysics, Vol. II} (University Science
  Books)

\bibitem[{{Shu} {et~al.}(1988){Shu}, {Lizano}, {Ruden}, \& {Najita}}]{shu88}
{Shu}, F.~H., {Lizano}, S., {Ruden}, S.~P., \& {Najita}, J. 1988, \apjl, 328,
  L19

\bibitem[{{Smith}(2014)}]{smith14}
{Smith}, N. 2014, \araa, 52, 487

\bibitem[{{Sridhar} \& {Goldreich}(1994)}]{sridhar94}
{Sridhar}, S., \& {Goldreich}, P. 1994, \apj, 432, 612

\bibitem[{{Stone} {et~al.}(1998){Stone}, {Ostriker}, \& {Gammie}}]{stone98}
{Stone}, J.~M., {Ostriker}, E.~C., \& {Gammie}, C.~F. 1998, \apjl, 508, L99

\bibitem[{{Swift} \& {Welch}(2008)}]{swift08}
{Swift}, J.~J., \& {Welch}, W.~J. 2008, \apjs, 174, 202

\bibitem[{{Tafalla} {et~al.}(2004){Tafalla}, {Myers}, {Caselli}, \&
  {Walmsley}}]{tafalla04a}
{Tafalla}, M., {Myers}, P.~C., {Caselli}, P., \& {Walmsley}, C.~M. 2004, \aap,
  416, 191

\bibitem[{{Truelove} {et~al.}(1997){Truelove}, {Klein}, {McKee}, {Holliman},
  {Howell}, \& {Greenough}}]{truelove97}
{Truelove}, J.~K., {Klein}, R.~I., {McKee}, C.~F., {et~al.} 1997, \apjl, 489,
  L179+

\bibitem[{{Turk} {et~al.}(2011){Turk}, {Smith}, {Oishi}, {Skory}, {Skillman},
  {Abel}, \& {Norman}}]{turk11}
{Turk}, M.~J., {Smith}, B.~D., {Oishi}, J.~S., {et~al.} 2011, \apjs, 192, 9

\bibitem[{{Vink} {et~al.}(2001){Vink}, {de Koter}, \& {Lamers}}]{vink01}
{Vink}, J.~S., {de Koter}, A., \& {Lamers}, H.~J.~G.~L.~M. 2001, \aap, 369, 574

\bibitem[{{Wallerstein} {et~al.}(1997){Wallerstein}, {Iben}, {Parker},
  {Boesgaard}, {Hale}, {Champagne}, {Barnes}, {K{\"a}ppeler}, {Smith},
  {Hoffman}, {Timmes}, {Sneden}, {Boyd}, {Meyer}, \& {Lambert}}]{wallerstein97}
{Wallerstein}, G., {Iben}, Jr., I., {Parker}, P., {et~al.} 1997, Reviews of
  Modern Physics, 69, 995

\bibitem[{{Wang} {et~al.}(2010){Wang}, {Li}, {Abel}, \& {Nakamura}}]{wang10}
{Wang}, P., {Li}, Z.-Y., {Abel}, T., \& {Nakamura}, F. 2010, \apj, 709, 27

\bibitem[{{Wilking} {et~al.}(2008){Wilking}, {Gagn{\'e}}, \&
  {Allen}}]{wilking08}
{Wilking}, B.~A., {Gagn{\'e}}, M., \& {Allen}, L.~E. 2008, {Star Formation in
  the {$\rho$} Ophiuchi Molecular Cloud}, ed. B.~{Reipurth}, 351

\end{thebibliography}
\bibliographystyle{apj}

\end{document}